\documentclass[review, 3p, table, xcdraw, authoryear]{elsarticle}

\usepackage{setspace}
\journal{Journal of \LaTeX\ Templates}
\makeatletter
\def\ps@pprintTitle{%
    \let\@oddhead\@empty
    \let\@evenhead\@empty
    \def\@oddfoot{\footnotesize\itshape
         {Preprint} \hfill\today}%
    \let\@evenfoot\@oddfoot
    }
\makeatother

\usepackage{doi}
\usepackage{booktabs}
\usepackage{multirow}
\usepackage{amssymb}
\usepackage{amsmath}
\usepackage{caption}

\usepackage{hyperref}
\usepackage{hyphenat}
\usepackage{orcidlink}
\usepackage{mathtools}
\usepackage{bm}

\usepackage{hyperref}
\hypersetup{
  colorlinks   = true, 
  urlcolor     = blue, 
  linkcolor    = blue, 
  citecolor   = blue 
}

\newcommand{\method}{MobilityGen} 

\newcommand{\beginsupplement}{%
        \setcounter{table}{0}
        \renewcommand{\thetable}{S\arabic{table}}%
        \setcounter{figure}{0}
        \renewcommand{\thefigure}{S\arabic{figure}}%
        \renewcommand{\thesection}{\Alph{section}}
     }

\begin{document}

\begin{frontmatter}

  \title{Deep Generative Model for Human Mobility Behavior}

  \author[ikg,keg]{Ye Hong\corref{correspondingauthor}}
  \cortext[correspondingauthor]{Corresponding author}
  \ead{ye.hong@keg.lu.se}

  \author[frs,ucl]{Yatao Zhang}
  \ead{yatao.zhang@ucl.ac.uk}

  \author[igp]{Konrad Schindler}
  \ead{schindler@ethz.ch}

  \author[ikg,frs]{Martin Raubal}
  \ead{mraubal@ethz.ch}

\address[ikg]{Institute of Cartography and Geoinformation, ETH Zurich}
\address[keg]{Department of Human Geography, Lund University}
\address[frs]{Future Resilient Systems, Singapore-ETH Centre, ETH Zurich}
\address[igp]{Photogrammetry and Remote Sensing, ETH Zurich}
\address[ucl]{Department of Geography, University College London}

  \begin{abstract}
    Understanding and modeling human mobility is central to challenges in transport planning, sustainable urban design, and public health.
    Despite decades of effort, simulating individual mobility remains challenging because of its complex, context-dependent, and exploratory nature. 
    Here, building on the activity-based view of daily mobility, we propose \method{}, a diffusion-based generative framework for simulating multi-attribute activity-travel sequences over days to weeks at large spatial scales.
    By linking behavioral attributes with environmental context, \method{} reproduces key patterns such as scaling laws for location visits, activity time allocation, and the coupled evolution of travel mode and destination choices.
    It reflects spatio-temporal variability and generates diverse and plausible mobility patterns consistent with the built environment.
    Beyond standard validation, \method{} enables analyses that have been difficult with earlier models, including how access to urban space varies across travel modes and how co-presence dynamics shape social exposure and segregation.
    Together, these results support an integrated, data-driven basis for fine-grained studies of human mobility behavior and its societal implications.
  \end{abstract}

  \begin{keyword}
    human mobility; activity schedules; deep generative model;  transport planning; travel behavior modeling
  \end{keyword}

\end{frontmatter}


\section{Introduction}
Human mobility is inherently complex and often appears unstructured.
From day to day, individuals visit locations across various spatial scales, using multiple transport modes at different periods of the day to engage in a range of activities.
Indeed, movement decisions involve intricate psychological planning~\citep{axhausen_activitybased_1992} and are influenced by individual factors such as physical ability~\citep{burbidge2009active} and socioeconomic status~\citep{hanson1981travel}.
Despite this complexity, human mobility is fundamentally constrained by physical infrastructures~\citep{jiang2009characterizing}, such as road and rail networks, and heavily influenced by the surrounding environment~\citep{handy2002built, hong2014built}. 
Ample empirical evidence demonstrates that mobility patterns follow simple scaling laws~\citep{brockmann_scaling_2006, gonzalez_understanding_2008, alessandretti_evidence_2018, schlapfer_universal_2021}, and exhibit striking regularities, for instance regarding spatial location choices~\citep{song_limits_2010}.
These collective regularities form the foundation for the realistic simulation of human mobility behaviors.
For transport planning, the key challenge is to simulate not only \emph{where} people travel, but also \emph{when} activities occur, and \emph{how} trips are made, while preserving their interdependencies across days.

Despite significant efforts and the importance of the task, simulating mobility behavior remains challenging.
Current approaches to synthesizing disaggregated travel demand, such as the Equasim pipeline~\citep{horl_synthetic_2021}, the OASIS framework~\citep{pougala_oasis_2023}, and various other methods based on the principle of utility maximization~\citep{bhat2000comprehensive, auld2012activity}, focus on activity participation and temporal constraints, while often simplifying the spatial structure of activity locations and travel.
%
%
Conversely, state-of-the-art mechanistic models, including the Container model~\citep{alessandretti_scales_2020}, the exploration and preferential return (EPR) model~\citep{song_modelling_2010}, and more recent variants derived from them~\citep{pappalardo_returners_2015, barbosa_effect_2015, zhao_characteristics_2021}, faithfully reproduce scaling laws of location visits, yet their limited modeling capacity restricts them from capturing the interactions between individuals' spatial preferences and their other movement decisions~\citep{pappalardo_future_2023}.
Neglecting those interactions can lead to misconceptions about individual space usage.
This has far-reaching consequences for downstream applications that depend on accurate spatial modeling, such as sustainable transportation~\citep{xu_planning_2018}, human-centered urban design~\citep{zhang_towards_2023}, and epidemic prevention~\citep{barreras_exciting_2024}. 

Data-driven generative models offer a possible route to simulation that preserves multi-attribute travel behavior at scale.
In recent years, learning-based models have revolutionized image and text processing.
Notably, natural language processing has experienced remarkable progress through new sequence processing techniques that excel at capturing complex patterns in large text collections, in particular, the transformer architecture~\citep{Vaswani_2017} and the denoising diffusion probabilistic model (DDPM)~\citep{sohl_deep_2015, ho_denoising_2020}.
Their effectiveness in modeling language has facilitated applications across diverse domains concerned with spatial or temporal sequences, such as protein structures~\citep{liu_novo_2024}, turbulent flows~\citep{li_synthetic_2024}, or human life events~\citep{savcisens_using_2024}.
Inspired by these advances, recent work has begun to adapt deep generative frameworks to human mobility and trajectory synthesis~\citep{chu_simulating_2024, songyiwen_controllable_2024, long_universal_2025}.
These studies highlight the promise of data-driven generation, while leaving open the challenge of jointly modeling the interdependent spatio-temporal and modal structure of daily mobility behavior at scale.
Following activity-based modeling and recent data-driven mobility prediction studies~\citep{axhausen_activitybased_1992, mo_individual_2022}, individual mobility can be represented as a sequence of discrete, time-ordered activity-travel events, each linking concurrent spatial, temporal, and modal attributes.
Still, language models are not directly applicable to human mobility behavior.
Generalizing deep generative modeling techniques, designed for one-dimensional word sequences, to multifaceted, context-dependent mobility behavior requires careful consideration.

Here, we introduce \method{}, which operationalizes this event-based view within a diffusion-based generative framework for daily mobility sequences unfolding over days and weeks at large spatial scales.
In this framework, where people go, when activities occur, and how they travel are modeled as coupled behavioral decisions rather than separable components.
We use a DDPM to learn the joint structure of multi-attribute event sequences from data.
Each event comprises multiple behavioral attributes, modeled as distinct but interdependent learning objectives, with their interactions captured through a shared latent representation that encodes meaningful structure across mode, time, and location. 
In addition, \method{} incorporates contextual features of the built environment, enabling the generation of diverse behavioral variations consistent with observed constraints and context.
As our experiments show, \method{} preserves national-scale visitation and transition regularities.
It also captures behavioral dependencies that span extended temporal horizons, producing synthetic sequences aligned with real-world patterns.
%
Together, these results demonstrate the utility of diffusion-based generative modeling for simulating coupled daily behavior at scale.
This enables the generation of comprehensive, fine-grained mobility data and supports applications ranging from testing sustainable transport strategies and assessing accessibility in urban design to analyzing co-presence dynamics that influence social segregation and epidemic response.

\section{Methodology}\label{sec:method}

\subsection{Event-level mobility representation and generative modeling}~\label{sec:modeling}
\begin{figure*}[!t]
  \centering
  \includegraphics[width=0.99\textwidth]{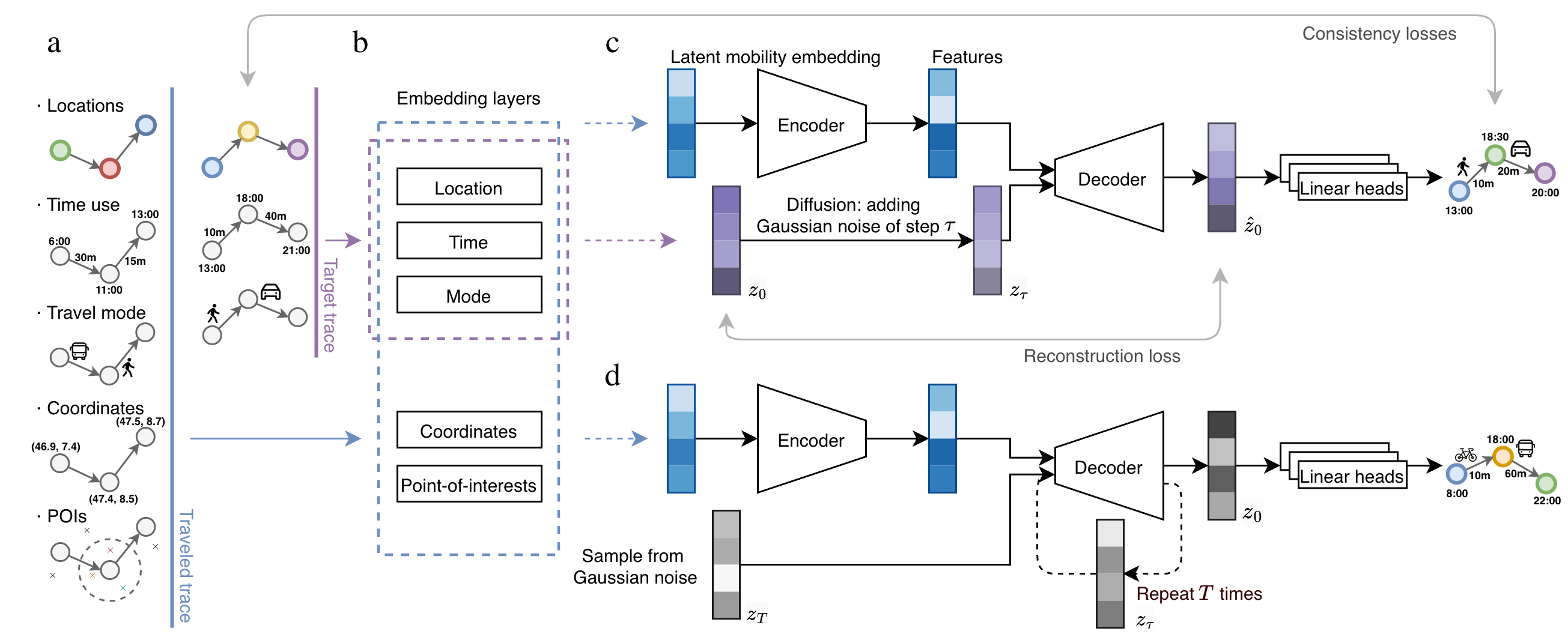}
  \caption{Modeling individual mobility with an event-level generative framework.
  \textbf{a}. We represent individual mobility as a sequence of chronologically ordered activity events, where each event has associated activity attributes (location, time use, and travel mode) and contextual information (coordinates and nearby POIs).
  \textbf{b}. Raw activity events are encoded into latent mobility embeddings through dedicated embedding modules.
  \textbf{c}. During training, the diffusion process adds Gaussian noise to the target sequence embedding $\mathbf{z}_0$. A transformer-based decoder learns to reverse this process and reconstruct a denoised embedding $\hat{\mathbf{z}}_0$. Features from the traveled sequence are extracted by an encoder and provided as guidance to the decoder. Final activity attributes are predicted via linear output heads. The model is trained by minimizing reconstruction and consistency losses between the generated and original sequences (see Methods).
  \textbf{d}. During inference, we sample a Gaussian noise vector $\mathbf{z}_T \sim \mathcal{N}(0, \mathbf{I})$ and iteratively refine it using the decoder, guided by encoder-extracted features. The denoised embedding $\mathbf{z}_0$ is then decoded into the sequence of activity events.
  Travel mode icons adapted from Flaticon.com.}
  \label{fig:overview}
\end{figure*}

To characterize individual mobility and support an event-level model, we analyzed global navigation satellite system (GNSS) trajectories from a smartphone-based travel survey that recorded the mobility behavior of thousands of participants across Switzerland. 
%
%
GNSS traces are segmented into stationary periods, i.e., fixed locations where individuals perform activities, and movement sections corresponding to transits between these activity locations. 
We then calculate activity attributes (such as time use and travel mode) and project the activity locations onto a hierarchical multi-resolution grid, with the highest resolution being $\approx$500$\times$500m\textsuperscript{2} per cell. 
%
Following activity-based modeling and prior data-driven studies~\citep{axhausen_activitybased_1992, mo_individual_2022}, we represent human mobility as sequences of activity events, each linked to specific attributes such as location, time use, and travel mode, along with contextual information (Figure~\ref{fig:overview}a). 
We particularly focus on contexts that characterize the built environment, including the coordinate geometry of locations and points of interest (POIs) that represent urban functional facilities.

The modeling challenge is therefore to learn how these heterogeneous event attributes vary together over time and across contexts.
Traditional methods often focus on a single aspect, such as location choices~\citep{alessandretti_scales_2020} or travel mode preferences~\citep{reck_mode_2022}, or are limited to describing simple behavioral dependencies~\citep{jiang_timegeo_2016, horl_synthetic_2021}.
\method{} addresses this gap by casting the joint spatial, temporal, and modal structure of activity-travel events as a generative modeling problem, learned with a diffusion-based framework conditioned on contextual information.

\method{} employs learnable embedding modules to map raw attributes and contexts to a shared embedding space (Figure~\ref{fig:overview}b). 
Learning comprehensive mobility embeddings enables the model to encode the rich multi-attribute dependencies of activity events.
We utilize a DDPM~\citep{sohl_deep_2015, ho_denoising_2020}, a powerful scheme for learning complex, high-dimensional distributions, to capture the structure of mobility event sequences.
In a DDPM, the original sequence is gradually corrupted with tractable random noise, and an encoder-decoder network (in our case, a transformer~\citep{Vaswani_2017}) learns the reverse denoising process (Figure~\ref{fig:overview}c). 
Once trained, the (reverse) process can generate new, plausible sequences from Gaussian noise (Figure~\ref{fig:overview}d). 
Additionally, a transformer encoder derives features from the observed travel sequence, acting as a guiding signal during denoising for generating event embeddings.
Finally, linear decoder heads extract the respective activity attributes from the predicted embeddings.
Together, these attributes form the final output, namely a new, simulated sequence of activity events for an individual, sampled in a way that respects observed transition and combination patterns across attributes.

\subsection{Problem definition}

Formally, we focus on conditional mobility generation, aiming to simulate a sequence of future activity events $ \mathbf{y}=\left [ y_1, y_2,...,y_{k_1} \right ]$ based on a preceding sequence of observed events $\mathbf{x}=\left [ x_1, x_2,...,x_{k_2} \right ]$ by the same individual. 
Each activity event $x_i$ or $y_i$ is represented as a tuple $\langle l, t, d, m \rangle$, where $l$ denotes the location identifier, $t$ the start time, $d$ the duration, and $m$ the travel mode. To incorporate spatial context, each event in the observed sequence $\mathbf{x}$ is further associated with geographic coordinates $g$ and land-use context $c$ describing the surrounding environment.

\subsection{Diffusion-based modeling of discrete activity sequences}

Diffusion models, grounded in non-equilibrium thermodynamics~\citep{sohl_deep_2015}, are a class of likelihood-based deep generative models that excel at learning true data distributions and generating new data samples. 
These models have overcome several limitations of earlier approaches, with recent advancements in reparameterization~\citep{ho_denoising_2020} and inference efficiency~\citep{lu2022dpm} leading to breakthroughs in handling continuous data distributions, such as those in image and audio generation~\citep{yang_diffusion_2023}.
This progress has motivated researchers to extend the framework to discrete data, such as text, resulting in the development of discrete diffusion models for categorical distributions~\citep{austin_structured_2021} and embedding diffusion models that incorporate additional discrete-to-continuous mapping steps~\citep{li_diffusion_lm_2022, gong_diffuseq_2023}.
Here, we build on the conditional embedding DDPM by~\citet{li_diffusion_lm_2022} and generalize the framework to accommodate multifaceted mobility behavior. 
This formulation suits mobility event sequences because it denoises the full target sequence jointly.
Joint denoising lets the model capture global sequence coherence, where location, start time, duration, and mode are mutually constrained across a day or multiple days. 
It also mitigates the train-inference mismatch of autoregressive rollouts, where reliance on ground-truth histories during training and on the model's own predictions at inference can compound errors over long event horizons.
Stochastic denoising further provides a natural mechanism for sampling multiple plausible continuations from the same observed history.

The employed diffusion model introduces two Markov chains: a forward chain that progressively destructs data by injecting noise and a reverse chain that reconstructs the original data from noised samples.
Formally, given a continuous data point $\mathbf{z}_0$ sampled from a real-world data distribution $q(\mathbf{z})$, i.e., $\mathbf{z}_0\sim q(\mathbf{z})$, the forward Markov chain gradually corrupts $\mathbf{z}_0$ into a standard Gaussian noise $\mathbf{z}_T\sim \mathcal{N}(0, \mathbf{I})$ in $\tau \in [1, 2, ..., T]$ steps:
\begin{equation}
    \label{equation:forward}
    q(\mathbf{z}_1,..., \mathbf{z}_{T}\mid \mathbf{z}_{0}) \coloneqq \prod_{\tau=1}^{T}q( \mathbf{z}_{\tau}\mid \mathbf{z}_{\tau-1})
\end{equation}
where $\mathbf{z}_1,..., \mathbf{z}_{T}$ are the chain of noisy latent variables. 
For each forward step, the perturbation is controlled by adding Gaussian noise with a predefined step-dependent mean $\mathbf{\mu}_\tau(\mathbf{z}_{\tau}) = \sqrt{1-\beta_\tau}\mathbf{z}_{\tau-1}$, and variance $\mathbf{\Sigma}_\tau(\mathbf{z}_{\tau}) = \beta_\tau\mathbf{I}$, with $\beta_\tau \in (0,1)$: 
\begin{equation}
    \label{equation:forward_step}
    q(\mathbf{z}_\tau\mid \mathbf{z}_{\tau-1}) \coloneqq \mathcal{N}(\mathbf{z}_\tau; \sqrt{1-\beta_\tau}\mathbf{z}_{\tau-1}, \beta_\tau\mathbf{I})
\end{equation}

Given the forward chain, the reverse denoising process aims to obtain $p_{\theta}(\mathbf{z}_{\tau-1}\mid \mathbf{z}_\tau)$, with which we can gradually reconstruct the original data $\mathbf{z}_0$ via sampling from $\mathbf{z}_T$ and apply through a Markov chain: 
\begin{equation}
    \label{equation:reverse}
    p_{\theta}(\mathbf{z}_0,..., \mathbf{z}_{T}) \coloneqq p(\mathbf{z}_T) \prod_{\tau=1}^{T}p_{\theta}( \mathbf{z}_{\tau-1}\mid \mathbf{z}_{\tau})
\end{equation}
where $p(\mathbf{z}_T)=\mathcal{N}(\mathbf{z}_T; 0, \mathbf{I})$. As the variance $\beta_\tau$ is, by definition, small, the reverse process follows the same functional form as the forward process, i.e., a Gaussian distribution~\citep{sohl_deep_2015}:
\begin{equation}
    \label{equation:reverse_step}
    p_{\theta}(\mathbf{z}_{\tau-1} \mid \mathbf{z}_\tau ) \coloneqq \mathcal{N}(\mathbf{z}_{\tau-1}; \mathbf{\mu}_{\theta}(\mathbf{z}_{\tau},\tau), \mathbf{\Sigma}_{\theta}(\mathbf{z}_{\tau},\tau))
\end{equation}
where the mean $\mathbf{\mu}_{\theta}(\mathbf{z}_{\tau},\tau)$ and variance $\mathbf{\Sigma}_{\theta}(\mathbf{z}_{\tau},\tau)$ are parameterized by deep neural networks with learnable parameters $\theta$. Training is performed by maximizing the marginal likelihood of the data $\mathbb{E}[\log p_{\theta}(\mathbf{z}_0)]$, and the canonical objective is the variational lower bound $\mathcal{L}_{vlb}$ of $\log p_{\theta}(\mathbf{z}_0)$ (for a detailed derivation of the objective, see~\citet{sohl_deep_2015}):
\begin{equation}
    \label{equation:loss_vlb}
    \mathcal{L}_{vlb}=\underset{q(\mathbf{z}_1,..., \mathbf{z}_{T}\mid \mathbf{z}_{0})}{\mathbb{E}} 
    \left[ 
    D_{\mathrm{KL}} \left( q(\mathbf{z}_T\mid \mathbf{z}_0) \parallel p(\mathbf{z}_T) \right) + \sum_{\tau>1}D_{\mathrm{KL}}(q(\mathbf{z}_{\tau-1}\mid \mathbf{z}_\tau,\mathbf{z}_0) \parallel p_{\theta}(\mathbf{z}_{\tau-1}\mid \mathbf{z}_{\tau}))-\log p_{\theta}(\mathbf{z}_{0}\mid \mathbf{z}_{1}) \right]
\end{equation}
where $D_{\mathrm{KL}}(\cdot \parallel \cdot)$ denotes the Kullback–Leibler (KL) divergence, and $q(\mathbf{z}_{\tau-1}\mid \mathbf{z}_\tau,\mathbf{z}_0)$ is the forward process posterior, which follows a Gaussian distribution with mean $\hat{\mathbf{\mu}}(\mathbf{z}_{\tau},\mathbf{z}_{0})$ and variance $\hat{\beta}\mathbf{I}$ that can be calculated from $\mathbf{z}_{0}$, $\mathbf{z}_{\tau}$ and $\beta_\tau$. The objective can be further simplified by reweighting each KL divergence term to obtain a mean-squared error reconstruction loss: 
\begin{equation}
    \label{equation:loss_simple}
    \mathcal{L}_{reconstruct}=\underset{\mathbf{z}_0,\mathbf{z}_\tau,\tau}{\mathbb{E}} \left\| \hat{\mathbf{z}}_{{\theta}}(\mathbf{z}_{\tau},\tau) - \mathbf{z}_{0} \right\|^2
\end{equation}
Here, $\hat{\mathbf{z}}_{\theta}(\mathbf{z}_{\tau},\tau)$ represents the prediction of the original data given $\mathbf{z}_{\tau}$ and $\tau$. Intuitively, this parameterization can be interpreted as training a neural network to predict the original ground truth data from an arbitrarily corrupted version.

To apply continuous diffusion to discrete data, embedding diffusion models introduce an additional embedding process in the forward process and a rounding step in the reverse process.
The embedding process transforms discrete tokens into a continuous latent space via a learnable embedding function, enabling the conventional diffusion process to be applied. Given the target sequence $ \mathbf{y}$, the input $\mathbf{z}_0$ is obtained in the embedding space as $\mathbf{z}_0\sim \mathcal{N}(\bm{e}_{\phi}(\mathbf{y}), \beta_0\mathbf{I})$, where $\bm{e}_{\phi}(\cdot)$ represents the embedding function parameterized by $\phi$. 
The rounding step translates the diffusion prediction in the embedding space back to discrete tokens: $p_{\phi}(\mathbf{y}\mid \mathbf{z}_0)$, whose parameter $\phi$ is shared with the embedding function. 
As a result, the neural network is trained end-to-end, optimizing both the reconstruction loss from the diffusion process and the consistency loss from the rounding step:
\begin{equation}
    \label{equation:loss_final}
    \mathcal{L} = \mathcal{L}_{reconstruct} + \underbrace{\underset{\mathbf{y},\mathbf{z}_0}{\mathbb{E}}\left [ -\log p_{\phi}(\mathbf{y}\mid \mathbf{z}_0) \right ] }_{\mathcal{L}_{consistent}}
\end{equation}

\subsection{Neural network architecture}
\method{}'s neural network comprises several components: an embedding module that maps activity attributes into an embedding space, an encoder that extracts information from traveled traces, a decoder that reconstructs target sequences from their corrupted versions, and task-specific linear heads that ensure consistency mapping.

The embedding process converts raw activity attributes of various formats into latent space representations.
For categorical attributes with discrete values, such as the location identifier $l$ and travel mode $m$, we introduce lookup matrices $\mathcal{E}_l$ and $\mathcal{E}_m$ that map attributes to the embedding space, $\mathcal{E}_l: \mathcal{V}_l\to \mathbb{R}^{d_{emb}}$ and $\mathcal{E}_m: \mathcal{V}_m\to \mathbb{R}^{d_{emb}}$, where $\mathcal{V}$ represents the set of possible values and $d_{emb}$ denotes the number of hidden dimensions. Intuitively, each row in $\mathcal{E}$ corresponds to the representation of a specific value. 
%
For continuous attributes, such as the start time $t$ and duration $d$, we use small feed-forward networks $f_t$ and $f_d$ to map the value into the hidden space, $f_t: \mathbb{R}^1\to \mathbb{R}^{d_{emb}}$ and $f_d: \mathbb{R}^1\to \mathbb{R}^{d_{emb}}$. 
The representation for an event record $y_i$ in the target sequence is obtained by linearly combining the above-introduced transformations:
\begin{equation}
    \label{equation:attribute_embed}
    \bm{e}_{\phi}(y_i) = \mathcal{E}_l(l_i) + \mathcal{E}_m(m_i) + f_{t}(t_i) + f_d(d_i)
\end{equation}

For each record $x_k$ in the traveled sequence, we also incorporate the geometry \( g_k \) and surrounding land-use functions \( c_k \). 
The geometry is modeled using the Space2vec method~\citep{mai_multi_2020}, which transforms raw coordinates in a projected coordinate system into vector representations via sine and cosine base functions. 
Land-use functions are captured using the location-POI embedding method proposed by~\citet{Hong_context_2023}, which considers POIs from various categories within a location and applies Latent Dirichlet Allocation (LDA) to derive their functional descriptions. 
The relationships between these movement-related contexts and the activity embeddings are learned through feed-forward networks $f$ and linearly combined via a residual connection:
\begin{align}
    \label{equation:context_embed}
    \bm{e}_{\phi}(x_k)  & = \mathbf{h}(x_k) + f_{coord}\left( s2v(g_k) \oplus \mathbf{h}(x_k)\right) + f_{poi}\left(lda(c_k) \oplus \mathbf{h}(x_k)\right)    \\
    \text{with} \quad  \mathbf{h}(x_k) & = \mathcal{E}_l(l_k) + \mathcal{E}_m(m_k) + f_{t}(t_k) + f_d(d_k)
\end{align}
where $\oplus$ denotes the concatenation operation. The final traveled sequence representation $\bm{e}_{\phi}(\mathbf{x})$ and target sequence representation $\bm{e}_{\phi}(\mathbf{y})$ consist of concatenated activity records in their sequential order. The parameters involved in the embedding process are optimized during network training.

We utilize the transformer encoder-decoder architecture proposed by~\citet{Vaswani_2017} as the network backbone. The encoder processes the traveled sequence representation through \( L \) encoder blocks, each with an identical architecture consisting of a multi-head self-attention layer and a feed-forward layer with two linear operations separated by a non-linear activation function. Residual connections~\citep{He_16}, layer normalizations~\citep{ba_layer_2016}, and dropout~\citep{hinton2012improvingneuralnetworkspreventing} are applied to each layer. The decoder, which processes the corrupted target sequence representation, contains $L$ decoder blocks structured similarly to the encoder. In addition to the two layers in each encoder block, the decoder incorporates an intermediate layer that performs multi-head attention over the encoder stack's output.

The multi-head attention layer applies $H$ scaled dot-product attention operations in parallel. 
Input representations have model dimension $d_{model}$. 
Within each head, queries and keys have dimension $d_k$ and values have dimension $d_v$, with $d_k = d_v = d_{model}/H$. 
Following \citet{Vaswani_2017}, scaled dot-product attention within each head is computed after the head-specific projections as:
  \begin{equation}
    \text{Attention}(\mathbf{Q}, \mathbf{K}, \mathbf{V}) 
    = \text{softmax}\!\left(\frac{\mathbf{Q}\mathbf{K}^T}{\sqrt{d_k}}\right)\cdot \mathbf{V}
  \end{equation}

Multi-head attention linearly projects the input matrices $\mathbf{Q}$, $\mathbf{K}$, $\mathbf{V}$ from $d_{model}$ into $H$ head-specific subspaces, applies attention within each head, and concatenates the results:
  \begin{align}
    \text{MultiHead}(\mathbf{Q}, \mathbf{K}, \mathbf{V}) & = (head_1 \oplus \cdots \oplus head_H)\cdot \mathbf{W}^O \\
    \text{with} \quad head_i & = \text{Attention}(\mathbf{Q}\mathbf{W}^Q_i, \mathbf{K}\mathbf{W}^K_i, \mathbf{V}\mathbf{W}^V_i)
  \end{align}
where $\mathbf{W}^Q_i, \mathbf{W}^K_i \in \mathbb{R}^{d_{model} \times d_k}$, $\mathbf{W}^V_i \in \mathbb{R}^{d_{model} \times d_v}$, and $\mathbf{W}^O \in \mathbb{R}^{H d_v \times d_{model}}$ are learned during training. In the decoder's multi-head attention layers, the queries are derived from the previous decoder block, while the keys and values come from the encoder stack's output. In the self-attention layers of both the encoder and decoder, the keys, values, and queries are all set to the output of the previous block. Additionally, causal masks are applied in the decoder's self-attention operations to prevent access to information from ``future'' sequence positions, i.e., the event at position $k$ can only attend to information from position $k$ and earlier. Ultimately, the encoder extracts the feature representation $x$ from the traveled mobility sequence, which is then used in the decoder to reconstruct the target trace representation $\hat{\mathbf{z}}_{\theta}(\mathbf{z}_{\tau},\tau, x)$.

Since we aim to generate new multifaceted activity events, the neural network must include multiple task-specific heads in the rounding step to ``translate'' from the embedding space into respective activity attributes. We introduce linear feed-forward layers $g_l: \mathbb{R}^{d_{emb}} \to \mathbb{R}^{\left| \mathcal{V}_l\right|}$, $g_m: \mathbb{R}^{d_{emb}} \to \mathbb{R}^{\left|\mathcal{V}_m\right|}$, $g_t: \mathbb{R}^{d_{emb}} \to \mathbb{R}^1$, and $g_d: \mathbb{R}^{d_{emb}} \to \mathbb{R}^1$ for directly mapping to location identifier, travel mode, start time, and duration, respectively. 

The parameters of $g_l$ and $g_m$ are tied to the corresponding embedding modules, and their outputs are processed with a softmax operation to ensure a valid probability distribution over the set of possible values:
  \begin{equation}
    P(\hat{\mathbf{l}}) = \text{softmax} \left(g_l(\mathbf{z}_{0} ) \right) \qquad P(\hat{\mathbf{m}}) = \text{softmax} \left(g_m(\mathbf{z}_{0}) \right) 
  \end{equation}
where $P(\hat{\mathbf{l}}) \in \mathbb{R}^{k_1 \times \left| \mathcal{V}_l\right|}$ and $P(\hat{\mathbf{m}}) \in \mathbb{R}^{k_1 \times \left| \mathcal{V}_m\right|}$ are matrices containing the predicted probabilities for all locations and travel modes, respectively. The rounding errors are measured as the average of multi-class cross-entropy loss across the sequence:
  \begin{equation}
    \mathcal{L}_{l} = -\frac{1}{k_1}\sum_{i=1}^{k_1}\sum_{j=1}^{\left| \mathcal{V}_l\right|}P(\mathbf{l})_{[i,j]}\log(P(\hat{\mathbf{l}})_{[i,j]}) \qquad 
    \mathcal{L}_{m} = -\frac{1}{k_1}\sum_{i=1}^{k_1}\sum_{j=1}^{\left| \mathcal{V}_m\right|}P(\mathbf{m})_{[i,j]}\log(P(\hat{\mathbf{m}})_{[i,j]})
  \end{equation}
where $P(\cdot)_{[i,j]}$ extracts the value at row $i$ and column $j$, and $P(\mathbf{l})$ and $P(\mathbf{m})$ are the one-hot represented ground truth, e.g., $P(\mathbf{l})_{[i,j]} = 1$ if the chosen location at step $i$ is the $j$\textsuperscript{th} location, and $P(\mathbf{l})_{[i,j]} = 0$ otherwise. 

The rounding losses for start time and duration are measured directly as the mean squared error after the linear heads:
  \begin{equation}
    \mathcal{L}_{t} = \left\| g_t(\mathbf{z}_{0}) - \mathbf{t} \right\|^2 \qquad 
    \mathcal{L}_{d} = \left\| g_d(\mathbf{z}_{0}) - \mathbf{d} \right\|^2
  \end{equation}

Finally, we obtain the consistency loss $\mathcal{L}_{consistent}$ by combining the task-specific rounding losses:
  \begin{equation}
    \label{equation:combine_loss}
    \mathcal{L}_{consistent} = \alpha_1\mathcal{L}_{l} + \alpha_2\mathcal{L}_{m} + \alpha_3\mathcal{L}_{t} + \alpha_4\mathcal{L}_{d} 
  \end{equation}
where the $\alpha$ values balance the contributions of the four attributes.
The reconstruction and consistency losses play complementary roles. 
The reconstruction loss in Eq.~\ref{equation:loss_simple} is applied to the joint event embedding $z_0$, which combines location, start time, duration, and mode information, and therefore supervises the full latent event representation directly.
The consistency losses in Eq.~\ref{equation:combine_loss} then act as attribute-level constraints, mapping the reconstructed embedding back to observed activity attributes.

\section{Experiment}

\subsection{Data sources and preprocessing}

\method{} was trained on a large-scale GNSS-based mobility dataset comprising high\hyp{}resolution individual trajectories collected across Switzerland.  
The dataset contains billions of GNSS tracking records from approximately 3,700 individuals between 2019 and 2022, and was produced as part of the \textit{Mobility Behaviour in Switzerland} (MOBIS) project, which investigates the behavioral impacts of mobility pricing~\citep{molloy_mobis_2022}.
Starting in autumn 2019, participants provided socio-demographic and travel-related information via a survey and recorded their movements over an eight-week period using a smartphone tracking app.
The app collected GNSS samples triggered by motion events detected through onboard sensors (e.g., accelerometer).
Following the main study, participants were invited to join the MOBIS-COVID extension, which assessed the impact of the COVID-19 pandemic on travel behavior. 
Roughly one-third of the original participants continued recording, with 730 individuals contributing data through the end of 2022.
For further details on the study design and findings, see~\citet{molloy_mobis_2022}.

We selected 2,113 individuals who had been tracked for at least 50 days and whose locations were known at least 60\% of the time.
This resulted in a dataset with an average tracking duration of 138 days and an average temporal coverage of 88\% across individuals.
The dataset provides substantial longitudinal depth and multi-attribute annotation across diverse Swiss contexts, supporting learning of the coupled spatio-temporal and modal structure of daily mobility.

Human mobility diaries and activity attributes were derived from raw GNSS tracking records using the standard movement data model~\citep{axhausen2007definition}.
The tracking app analyzed motion metrics such as speed and acceleration from built-in smartphone sensors to segment GNSS traces into movement \textit{stages} and stationary \textit{staypoints}.
Travel mode labels were inferred for each stage from the same motion data. During the tracking study, participants were encouraged to verify or correct the imputed mode labels and to annotate staypoints with their corresponding activity purposes.
We identified \textit{activity staypoints} as those where meaningful activities (e.g., home or restaurants) occurred, defined as stays longer than 25 minutes or carrying an activity label other than \textit{waiting}.
These staypoints were spatially aggregated into \textit{locations}, representing repeated visits to the same place, using the Trackintel framework~\citep{Martin_trackintel_2023} with parameters \(\epsilon = 20\) and \(num\_samples = 1\).
The resulting locations were projected onto Google’s S2Geometry grid system, a hierarchical tessellation of the Earth's surface.
We used a progressive gridding strategy from level 10 (coarse) to level 14 (fine, with a spatial resolution of approximately 500$\times$500m\textsuperscript{2}), selecting a finer level when more than two locations were projected onto the same cell.
This process yielded 142,575 location grids covering Switzerland, of which 28,741 were actively visited during the study period.

We define a \textit{trip} as the sequence of consecutive movement stages between two activity staypoints and assign its primary travel mode as the one used for the longest stage (one of car, train, bus, tram, bicycle, walk, or other).
In activity-based mobility analysis, travel is understood as a means to perform activities at different locations, making the activity itself the central unit of analysis~\citep{axhausen_activitybased_1992}.
Accordingly, we define each activity event with the following attributes: \textit{start time} as the departure time of the preceding trip, \textit{duration} as the sum of the trip and stay durations, and \textit{travel mode} as the primary mode used to reach the activity location. 
The preprocessed dataset consists of 1,079,918 chronologically ordered activity events, each associated with a start time, duration, and travel mode, and spatially projected onto 28,741 unique locations.

For acquiring the built environment information, we collected the POI dataset from OpenStreetMap (OSM) (\url{http://www.openstreetmap.org}), an open-source project that provides free and easily accessible digital map resources. 
Historical POI data for Switzerland were acquired from the end of 2022 to align with the time frame of the tracking dataset. 
The POI dataset was supplemented with OSM building data from the same year. For each building entry, we retained its attribute values but created a new POI by using the center point of its polygon geometry as its spatial representation.
After removing redundant and insignificant POIs, the final dataset consisted of 3,104,590 POIs, categorized into 19 first-level and 188 second-level groups. The second-level categories, representing functional type descriptions, were used to obtain urban functions for locations.

We divided the activity event sequences into non-overlapping training, validation, and test sets in a 7:2:1 ratio based on time, with the first 70\% of each individual's tracking days designated as training data, the middle 20\% as validation data, and the last 10\% as test data.
We constructed conditional sequence pairs by using mobility over the past three weeks as the traveled sequence and the subsequent activity events as the target sequence, requiring each target to span at least one calendar week.
Because individuals differ in activity frequency and observation coverage, the resulting sequences have variable lengths, denoted by \(k_2\) and \(k_1\) for the traveled and target sequences, respectively.
This process yielded 590,257, 104,010, and 36,285 traveled-target sequence pairs for training, validation, and testing, respectively.
The model is trained on these variable-length sequence pairs using sequence masking, allowing the learning process to reflect differences in individual mobility intensity.
Because this temporal split shares individuals between training and test, we additionally evaluate generalization to users entirely absent from training in Appendix~\ref{appendix:user_split}.

\subsection{Training procedure and computational cost}

We optimize the network parameters with Eq.~\ref{equation:loss_final} on the training set using the AdamW variant of stochastic gradient descent~\citep{loshchilov_decoupled_2018}, employing a weight decay of 0.01 and an adaptive learning rate that starts at $4e^{-4}$ and is progressively reduced to $2e^{-5}$ following a cosine schedule. 
Training is conducted over 250,000 steps, with a batch size of 64 traveled-target sequence pairs per step (approximately 32,000 tokens).
A linear learning rate warm-up~\citep{Vaswani_2017} is applied for the first 5,000 steps.
To stabilize training, we first optimize the network for 100,000 steps using only the $\mathcal{L}_{l}$ loss, setting $\alpha_1 = 1$, and $\alpha_2 = \alpha_3 = \alpha_4 = 0$.
For the remaining steps, we apply the efficient multiple-gradient descent algorithm for encoder-decoder architectures~\citep{sener_multi_task_2018} to dynamically adjust the $\alpha$ values, ensuring the parameter optimization process leads to a Pareto optimal solution, where no task can be improved without compromising at least one of the others.
We use the \textit{sqrt} noise schedule~\citep{li_diffusion_lm_2022} to determine the injected noise $\beta_\tau$ at each forward diffusion step and apply loss-weighted sampling to select the noise steps $\tau$ during training~\citep{gong_diffuseq_2023}. The transformer encoder and decoder each contain $L=6$ blocks and $H=8$ heads per attention layer. We set the forward diffusion steps to $T=2000$, latent embedding size to $d_{emb} = 128$, and the transformer model representation size to $d_{model} = 512$. These hyperparameters were determined through a random search, selecting the best-performing set that minimized the network loss on the validation set.

The complete training process takes $\approx$82 hours using distributed data parallelism across four NVIDIA 4090 GPUs with FP16 half-precision. 
During inference, the generation length $k_1$ is specified explicitly; for the main evaluation we set $k_1 = 50$ events, covering $12.7 \pm 4.0$ calendar days on average. 
The empirical target sequence is capped to the same length, ensuring that generated and observed sequences are compared over the same number of future activity events. 
This 50-event horizon is standardized across individuals and applied to all baseline sequence comparisons. 
Inference uses a down-sampled reverse process with $T=200$, and generating a new target sequence takes $50.4 \pm 4.1$ milliseconds on a single NVIDIA 3090 GPU. 
Appendix~\ref{appendix:sequence} reports sensitivity analyses for varying the input traveled sequence length $k_2$ and the generated target length $k_1$.

\subsection{Evaluation metrics}
We use metrics that summarize high-level mobility patterns at collective and individual levels to quantify the discrepancy between ground truth and generated activity sequences. We consider the following indicators:

\begin{itemize}
    \item Location visitation frequency~\citep{gonzalez_understanding_2008, song_modelling_2010}. Individuals' preferences in location choices are highly unbalanced and can be quantified by the relationship between the visitation frequency of locations $f_k$ and their rank $k$, which is reported to follow Zipf's law, i.e., $f_k \sim k^{-\xi }$, with $\xi \sim 1.2$.
    
    \item Evolution of radius of gyration~\citep{gonzalez_understanding_2008, alessandretti_scales_2020}. The radius of gyration $r_u$ measures the characteristic distance traveled by individual $u$ and is used to reflect mobility intensity in the spatial dimension. Its evolution after $k$ displacements $r_u(k)$ is defined as:
    \begin{equation}
        r_u(k) = \sqrt{\frac{1}{k}\sum_{i=1}^{k}(l^{(i)}-l^{cm}_u)^2}
    \end{equation}
    where $l^{(i)}$ represents the position of the $i$\textsuperscript{th} visited location, and ${l^{cm}_u = \frac{1}{k}\sum_{j=1}^{k}l^{(j)}}$ is the center of mass of the location sequence after $k$ displacements.
    
    \item Temporal mobility entropy~\citep{song_limits_2010}. Closely related to the theoretical predictability of location sequences, temporal mobility entropy $S_{temp}$ captures the full spatio-temporal order of location visits by measuring the probability $P(T_u^{'})$ of observing ordered subsequences $T_u^{'}$ within the location sequence $T_u$: 
    \begin{equation}
        S_{temp} = - \sum_{T_u^{'}\in T_u}P(T_u^{'})\log_2(P(T_u^{'}))
    \end{equation}
    We use the method described in~\citet{song_limits_2010} to estimate the temporal mobility entropy $S_{temp}$ for individuals.
    
    \item Mobility motifs~\citep{schneider_unravelling_2013}. Mobility motifs refer to recurring network patterns that emerge when daily location visits are abstracted as networks, where nodes represent activity locations and links denote transitions between them. We segmented the activity events by day and identified motif structures as those occurring in more than 0.5\% of all days. The construction of these motifs incorporates both spatial and temporal dimensions of activity participation. Mobility regularity is quantified by the presence and frequency of specific motifs in individuals' schedules~\citep{jiang_timegeo_2016}.
    
    \item Evolution of trip package choices~\citep{hong_conserved_2023}. The trip package concept was introduced to encapsulate the interactions within multifaceted activity-travel patterns. For each individual, a trip package is constructed when observing the usage of a travel mode to reach an activity location, i.e., a unique mode-location choice. Individuals continuously explore new activity-behavior options, and the number of unique trip packages observed over time quantifies the speed of this exploratory behavior.
\end{itemize}

\section{Results}\label{sec:result}

\subsection{Microscopic modeling reproduces location choice behavior} \label{sec:location}

To test whether \method{} preserves the core spatial regularities of daily mobility, we begin by evaluating its ability to capture location patterns, which constitute a basic validity check of the generation capability.
How individuals select spatial locations for different activities shapes their interactions with physical space and is a fundamental driver of many urbanization-induced challenges~\citep{moro_mobility_2021, nilforoshan_human_2023, liu2023beyond}.
Here, we focus on the simulated visits to locations and compare them with the real visits from observed sequences.

\begin{figure*}[!b]
  \centering
  \includegraphics[width=0.99\textwidth]{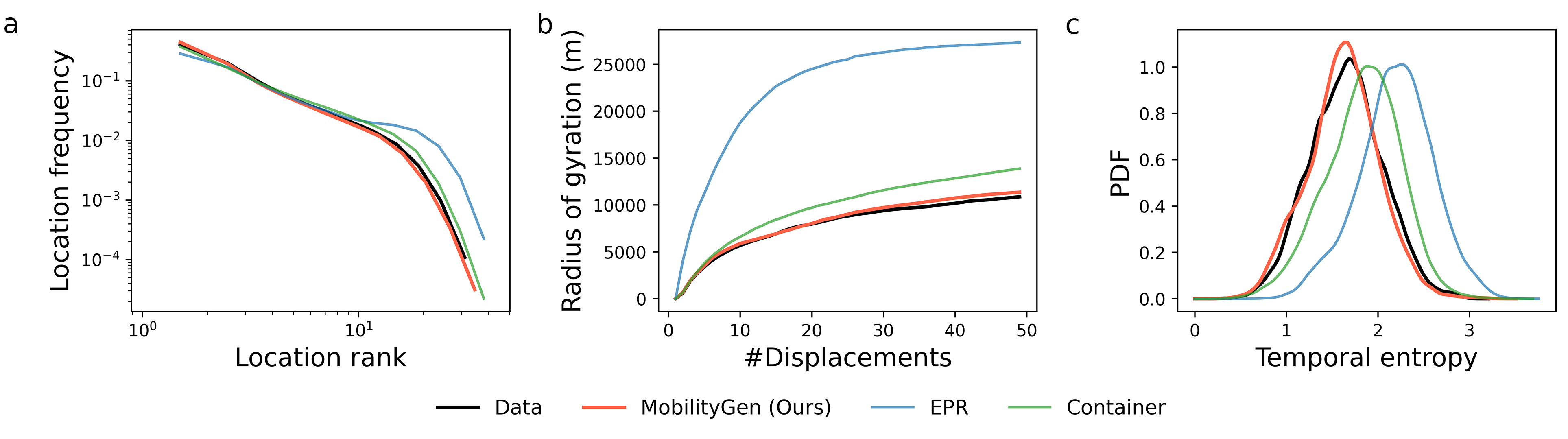}
  \caption{Evaluating microscopic mobility models with location metrics.
  \textbf{a}. Rank-frequency distribution of visited locations.
  \textbf{b}. Median radius of gyration as a function of the number of displacements.
  \textbf{c}. Distribution of temporal mobility entropy across individuals.
  All metrics are compared between real data (black), \method{} (red), EPR (blue), and Container (green). \method{} aligns most closely with empirical distributions across all metrics. }
  \label{fig:loc_metric}
\end{figure*}

We first assess the ability of \method{} to capture fundamental properties of location choices and compare its performance against two state-of-the-art microscopic mobility models, namely, exploration and preferential return (EPR)~\citep{song_modelling_2010}, and Container~\citep{alessandretti_scales_2020} (Figure~\ref{fig:loc_metric}). 
We begin with the average rank distribution of location visits (Figure~\ref{fig:loc_metric}a), a widely used measure of preference for activity locations~\citep{gonzalez_understanding_2008, barbosa_human_2018}. Our results indicate that \method{} provides a significantly better fit to real data than both the EPR and Container models, as confirmed by the likelihood ratio test~\citep{clauset2009power} (with $P < 10^{-3}$) and the Wasserstein distances~\citep{panaretos2019statistical} between the rank distributions (see Appendix~\ref{appendix:loc} and Table~\ref{tab:distance_loc}).
The radius of gyration, which measures the characteristic distance an individual covers~\citep{gonzalez_understanding_2008}, serves to quantify the movement intensity in space.
We observe that the median evolution of the radius of gyration across individuals, \( r_u(k) \), over displacement steps \( k \), is well described by logarithmic growth for both real and generated location sequences (Figure~\ref{fig:loc_metric}b), consistent with previous findings~\citep{alessandretti_scales_2020}.
However, when fitting \( r_u(k) \propto \alpha \log(k) \), only the value \( \alpha \) derived from \method{} agrees well with the real data (see Appendix~\ref{appendix:loc}).  
Finally, we evaluate the temporal mobility entropy that quantifies the predictability of location sequences, considering both the frequency and the order of visits~\citep{song_limits_2010}.
Using the likelihood ratio test and the Wasserstein metric, we further confirm that, compared against the EPR and Container models, \method{} most accurately approximates the true entropy distribution over the population (with $P < 10^{-3}$, Figure~\ref{fig:loc_metric}c and see Appendix~\ref{appendix:loc} and Table~\ref{tab:distance_loc}).
Additionally, we evaluated a representative set of state-of-the-art data-driven mobility models, including deep learning-based sequence prediction models and end-to-end generative models, and further location metrics (see Appendix~\ref{appendix:loc}).
Our findings indicate that \method{} consistently provides the best fit to real-world data. 
Evaluations with metrics from language modeling confirm that \method{} can simulate location visits that closely resemble the real sequence while avoiding regurgitation of input sequences and maintaining sufficient diversity (see Appendix~\ref{appendix:language}).
Finally, a held-out-individual analysis examines whether these location patterns generalize to users absent from training: visitation-frequency and uncorrelated-entropy patterns remain close to the seen-user reference under high source-location coverage, whereas measures of spatial extent show larger seen--unseen differences and deteriorate further when source-location coverage is lower (Appendix~\ref{appendix:user_split}).

\begin{figure*}[!t]
  \centering
  \includegraphics[width=0.99\textwidth]{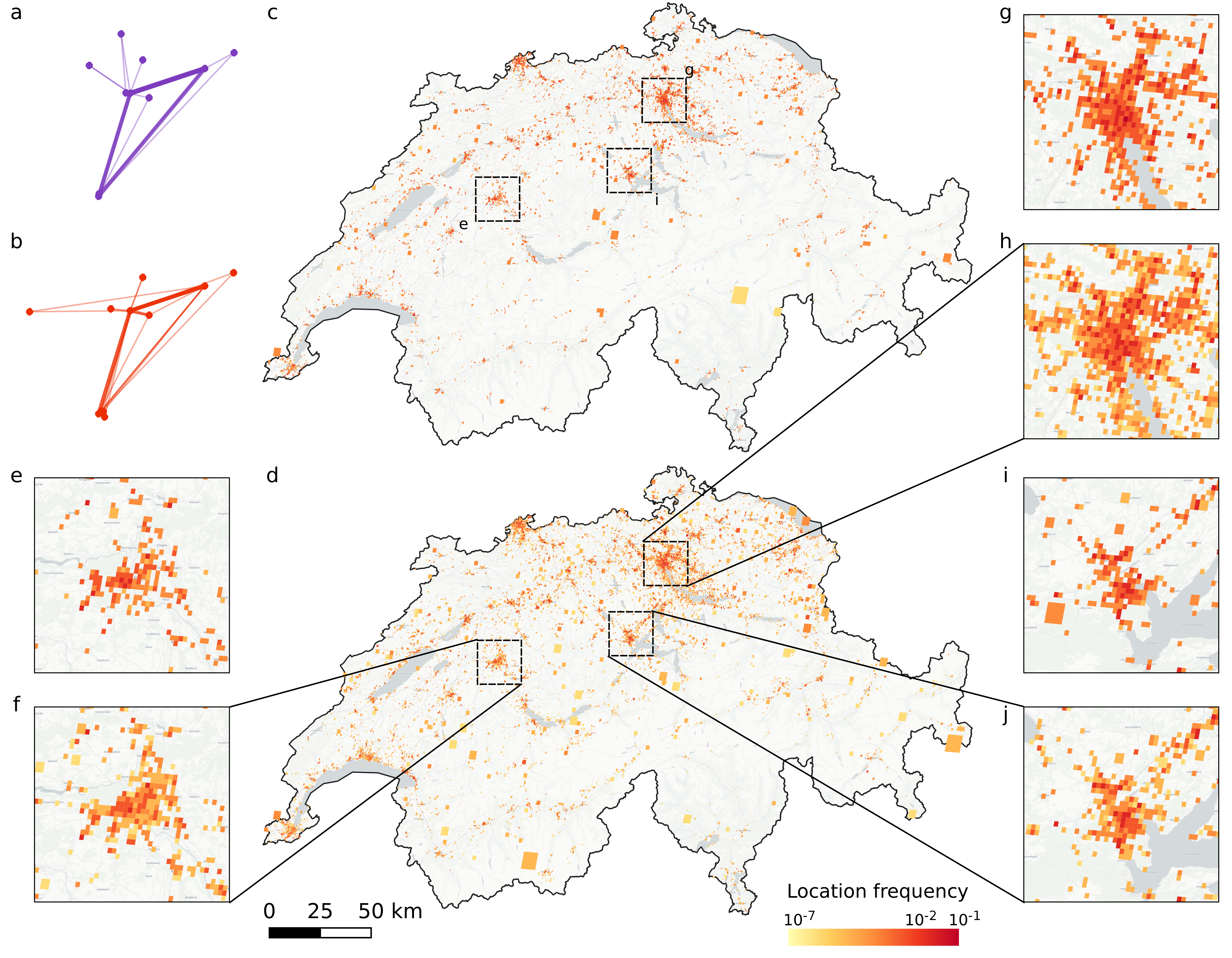}
  \caption{\method{} generates realistic spatial patterns of location visits across individual, urban, and national levels. We compare the spatial distribution of activity locations in real data (\textbf{a}, \textbf{c}, \textbf{e}, \textbf{g}, and \textbf{i}) with those generated by \method{} (\textbf{b}, \textbf{d}, \textbf{f}, \textbf{h}, and \textbf{j}). \textbf{a} and \textbf{b} show visits and transitions for a sample individual, where edge width is proportional to transition frequency. The generated sequence reflects routinely visited locations while also capturing exploratory behavior. \textbf{c} and \textbf{d} display aggregated visit frequencies across all individuals in Switzerland, with warmer colors indicating higher visitation intensity on a log scale. Panels \textbf{e}–\textbf{j} provide zoomed-in views for Bern (\textbf{e}, \textbf{f}), Zurich (\textbf{g}, \textbf{h}), and Lucerne (\textbf{i}, \textbf{j}). The generated sequences closely align with observed spatial patterns while exhibiting slightly greater diversity in the locations visited.}
  \label{fig:spatial_p}
\end{figure*}

To further evaluate the generated mobility patterns, we examine the spatial distribution of location visits. As an illustrative example, we map the observed and synthetic trajectories of a sample individual (Figure~\ref{fig:spatial_p}a and b). The comparison shows that \method{} effectively captures both visit and transition patterns at frequently visited locations (thick lines), while also generating occasional visits to less familiar locations (thin lines). Accurately modeling such exploratory behavior remains a central challenge in human mobility research~\citep{cuttone_understanding_2018}. 
To assess how well \method{} addresses this challenge at the population level, we compare the aggregated spatial distribution of location visits across Switzerland in the real and simulated data (Figure~\ref{fig:spatial_p}c and d). Visit frequencies are color-coded from low (yellow) to high (red). \method{} closely matches the concentration of frequently visited locations and also generates visits to less frequented areas, reflecting a balance between individual-level exploration and population-level spatial preference. Using the entropy $E=-\sum_{k} p(f_k) \log p(f_k)$ to quantify the dispersion of location visits $f_k$, we obtain 0.33 for the real data and 0.53 for the simulated data, indicating greater spatial dispersion in the latter.
To analyze spatial dynamics at a finer resolution, we examine three representative cities: Bern, Zurich, and Lucerne (Figure~\ref{fig:spatial_p}e-j), each characterized by distinct urban functions and mobility patterns. \method{} successfully reproduces key urban visitation structures in all three cases. However, spatial dispersion remains higher in the simulated data, with entropy values of 2.49 vs.\ 1.89 for Bern, 3.99 vs.\ 3.37 for Zurich, and 2.71 vs.\ 2.11 for Lucerne. Despite this discrepancy, \method{} compares favorably to other microscopic mobility models in replicating the location visit distributions (see Appendix~\ref{appendix:loc}, Table~\ref{tab:entropy_loc}). Complementary visual comparisons are provided at both national and urban levels in Appendix~\ref{appendix:loc}, Figure~\ref{suppl_fig:compare_full} and~\ref{suppl_fig:compare_detail}.

\subsection{Structured behavioral patterns across time, mode, and location} \label{sec:attribute}

\begin{figure*}[!t]
  \centering
  \includegraphics[width=0.99\textwidth]{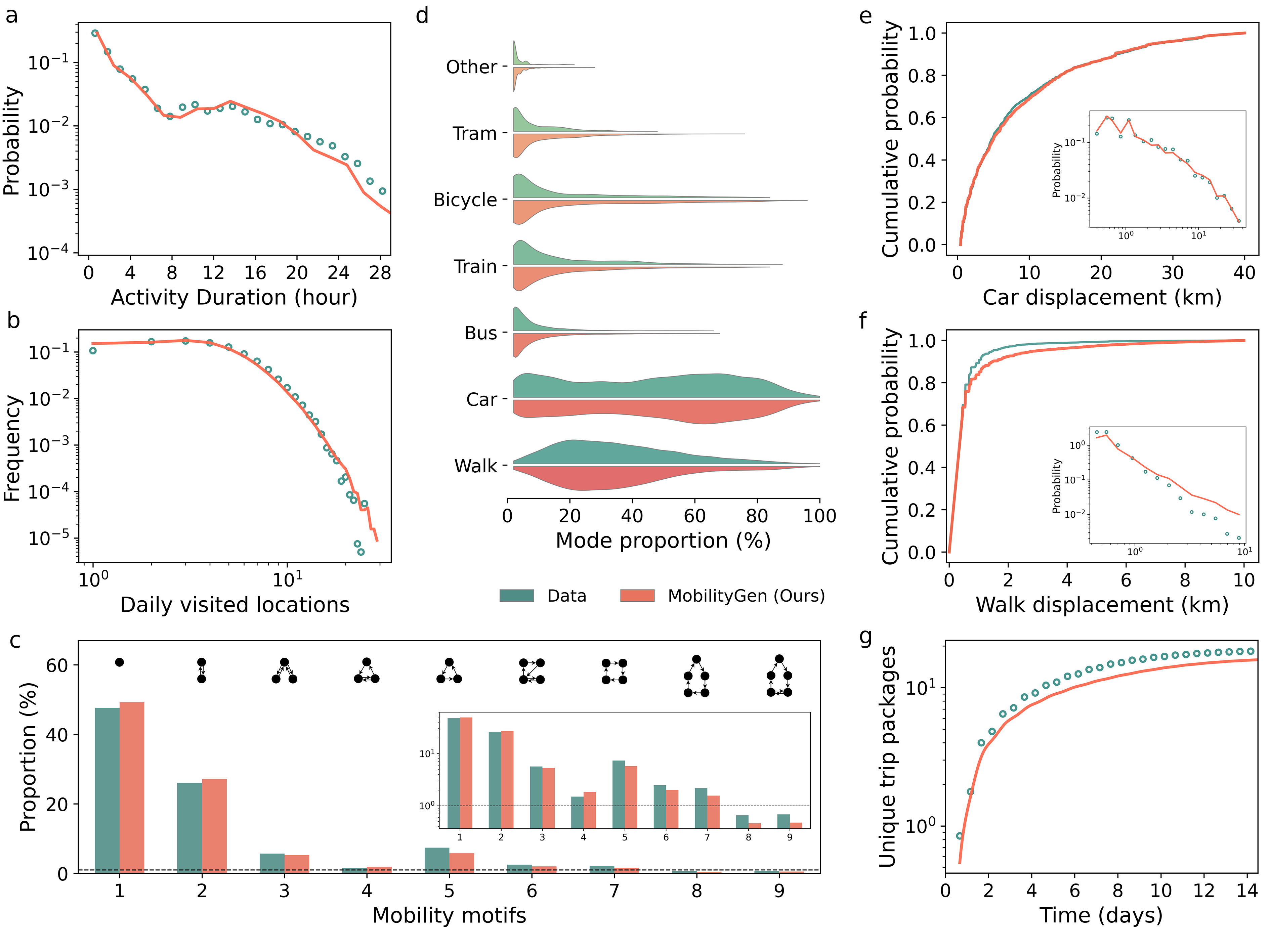}
  \caption{Generated activity events replicate individual activity-travel patterns across multiple dimensions.
  \textbf{a}. Distribution of activity durations.
  \textbf{b}. Distribution of daily visited locations.
  \textbf{c}. Distribution of mobility motifs, with motif diagrams shown above. The dashed line marks the 1\% threshold as a visual guide, and the inset presents the same distribution on a log scale.
  \textbf{d}. Violin plots of modal shares across individuals.
  \textbf{e}. Cumulative distribution of car displacements, with the log-log distribution shown in the inset.
  \textbf{f}. Cumulative distribution of walk displacements, with the log-log distribution shown in the inset.
  \textbf{g}. Evolution of unique trip packages over time, averaged across individuals.
  Across all dimensions, the activity events generated by \method{} (red) closely align with empirical patterns from real-world data (green), capturing key behavioral regularities and attribute interactions.}
  \label{fig:att_metric}
\end{figure*}

To examine the model beyond aggregate location patterns, we evaluate whether \method{} captures structured relationships among activity timing, travel mode, and location within unified activity events.
We assess these patterns for each sequence in the test set and compare cross-attribute regularities and interactions (Figure~\ref{fig:att_metric}).

Incorporating temporal information allows \method{} to reason about the scheduling of each activity. 
This temporal dimension is evident in the activity durations (Figure~\ref{fig:att_metric}a) and the number of daily visited locations (Figure~\ref{fig:att_metric}b), both of which closely match those observed in the real data.
To further examine patterns of activity participation, we construct mobility motifs~\citep{schneider_unravelling_2013}, which serve as high-level descriptors of regularities in activity location and scheduling.
%
%
%
\method{} accurately recovers the relative frequencies of both common and rare motifs, including simple two-location transitions and more complex patterns involving up to five locations (Figure~\ref{fig:att_metric}c).
We further benchmark \method{} against two state-of-the-art spatio-temporal mobility simulators, the Diary-based Trajectory Simulator (DITRAS)~\citep{pappalardo_data_2018} and TimeGeo~\citep{jiang_timegeo_2016}, using additional mobility metrics (see Appendix~\ref{appendix:attribute}).
Across all measures, \method{} more faithfully reproduces the interplay between activity timing and location choice, yielding the closest alignment with empirical behavior.
The held-out-individual analysis (Appendix~\ref{appendix:user_split}) further shows that this advantage persists for individuals absent from training.

Next, we incorporate user-verified travel mode labels into \method{} to jointly model mode choice in relation to activity events.
Figure~\ref{fig:att_metric}d compares the travel mode distributions from the observed and simulated activity sequences, with discrepancies quantified by the Wasserstein distance (see Appendix~\ref{appendix:attribute}, Table~\ref{tab:distance_all}).
Across all scenarios, \method{} reconstructs the mode preferences exhibited in real-world behavior.
Car and walking, the two most common modes, show excellent agreement between empirical and simulated shares.
Notably, \method{} captures subtler distinctions among public modes, including bus, train, and tram, accurately reflecting their relative frequencies.
We further assess mode-specific behavior by comparing the displacement distributions for car and walking trips (Figure~\ref{fig:att_metric}e and f).
For car travel, the simulated displacements closely align with the observed distribution (Kolmogorov–Smirnov statistic $KS=0.02$, with $P < 10^{-3}$). 
For walking, \method{} produces slightly longer displacements ($KS = 0.06$, $P < 10^{-3}$), potentially due to short trips being disproportionately affected by spatial discretization.
This effect could be mitigated in future work by adopting a finer location grid, although such refinement may increase computational complexity.

Finally, we assess the evolution of trip packages, defined as combinations of travel modes and activity locations~\citep{hong_conserved_2023}, to evaluate \method’s ability to model complex attribute interactions (Figure~\ref{fig:att_metric}g).
The average number of unique trip packages over time, denoted as \( Tp(t) \), follows a sub-linear growth over time \( t \), consistent with previous findings~\citep{hong_conserved_2023}. 
The scaling exponent \( \alpha \) in the relation \( Tp(t) \propto  t^{\alpha} \) is nearly identical between simulated ($\alpha = 0.85 \pm 0.01$) and empirical sequences ($\alpha = 0.85 \pm 0.01$), indicating that \method{} accurately captures the evolving diversity of travel mode and activity location choices over time.
Further ablation studies confirm that the full multi-attribute model achieves a more realistic representation of attribute interactions than simplified variants with fewer modeled attributes (see Appendix~\ref{appendix:ablation}).

To understand how \method{} captures attribute interactions, we analyze the structure of its learned embedding space. 
We apply densMAP~\citep{narayan_assessing_2021} to project the 128-dimensional embeddings into two dimensions for visualization (see Appendix~\ref{appendix:embedding}). 
The projections reveal that \method{} learns semantically meaningful representations: similar travel modes appear near each other, temporal attributes vary smoothly along continuous trajectories, and frequently visited locations are well-separated from infrequent ones. 
When multiple attributes are jointly embedded, the space further reveals distinct event clusters tied to specific locations, alongside a more diffuse region that likely reflects less location-dependent activities. 
Together, these patterns indicate that the embedding space offers a compact representation of multi-attribute mobility events and a natural basis for quantitative analysis of structure across mode, time, and location.
Notably, this structure emerges from the generative training objective alone, suggesting that \method{} encodes behavioral regularities and not merely the marginal output distributions.
By providing this shared representation, \method{} enables complex mobility dynamics to be inferred directly from behavioral sequences and facilitates diverse mobility analyses.

\subsection{Context enhances generalization to novel locations} \label{sec:context}

\begin{figure*}[!t]
  \centering
  \includegraphics[width=0.6\textwidth]{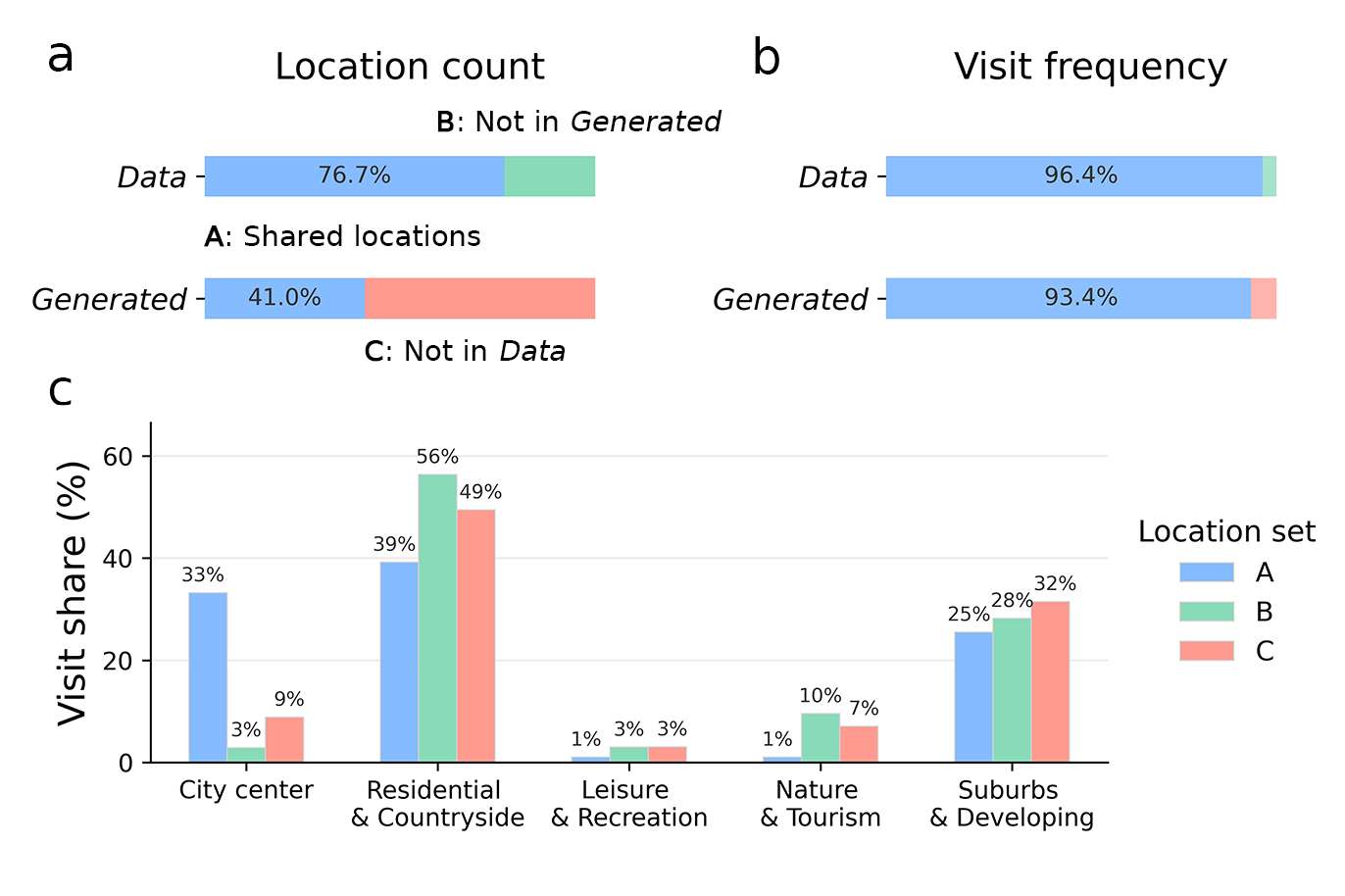}
  \caption{Point of Interest (POI) data supports the identification of locations associated with infrequent activities.
  \textbf{a}. We compare generated sequences with observed data, categorizing locations into those present in both (Set A), those found only in the data (Set B), and those found only in the generated sequences (Set C).
  \textbf{b}. As expected, locations in Set A are visited frequently, whereas those in Sets B and C are rarely visited.
  \textbf{c}. Relative visit frequencies by functional location type for each set. Functional types are inferred from POI distribution (see Methods). Sets B and C exhibit similar visit patterns, in contrast to the distribution observed for Set A.}
  \label{fig:poi_explore}
\end{figure*}

A key element of the proposed framework is that contextual information can be integrated when modeling individual travel decisions -- an aspect well-supported by the literature but not adequately covered by existing microscopic mobility models~\citep{pappalardo_future_2023}.
In \method, context features are combined with activity attributes through learnable representations, yielding a context-aware mobility model (see Figure~\ref{fig:overview}b).
We focus on spatial contexts that characterize the physical environment, namely coordinate geometry, which reveals movement directionality~\citep{zhao_unravelling_2024}; and POIs, which display urban functions and imply movement purposes~\citep{Hong_context_2023}. 
Further contexts could be integrated through appropriate representations; we defer this to future research.

We analyze location patterns to assess the influence of contextual factors on the simulation performance.
Generated locations are compared to observed data and categorized into three sets: Set A includes locations present in both sequences, while Sets B and C contain locations visited exclusively in the observed and generated sequences (Figure~\ref{fig:poi_explore}a).
Since \method{} replicates key location preferences (see Figure~\ref{fig:spatial_p}), Set A is expected to capture the locations where individuals conduct most of their activities.
Indeed, an analysis of the visit frequencies reveals that 96.4\% of the true visits and 93.4\% of the generated visits occur at locations in Set A (Figure~\ref{fig:poi_explore}b). 
In contrast, locations in Sets B and C receive relatively few visits, reflecting individuals' tendencies toward novelty-seeking and the exploration of unfamiliar locations~\citep{arentze_information_2005, song_modelling_2010, alessandretti_evidence_2018}.

To understand how \method{} models novelty-seeking behavior, we identify the functional types of locations using POI data.
Applying $k$-means clustering to POI representations of locations yields five distinct location classes, whose functional types are inferred from their spatial distributions and the enrichment of specific POI categories~\citep{yao_sensing_2017, niu_delineating_2021} (see Appendix~\ref{appendix:poi}).
This functional classification enables a detailed analysis of location preferences, revealing that visits in Set A predominantly occur in the \textit{City center}, whereas visits in Sets B and C are largely concentrated in \textit{Residential \& Countryside}.
Moreover, preferences for \textit{Leisure \& Recreation} and \textit{Nature \& Tourism} sites are markedly higher in Sets B and C than in Set A.
Using pairwise Wasserstein distances to quantify the similarity of visitation patterns across sets, we find that Sets B and C are closely aligned (Wasserstein distance: 0.11), while their distributions differ substantially from that of Set A (Set A vs. Set B: 0.57; Set A vs. Set C: 0.56).
In other words, \method{} captures key patterns of exploratory behavior, despite the inherent difficulty of characterizing novel locations and predicting which specific places individuals are likely to explore.

A comparison with a model that excludes contextual information confirms that incorporating context improves the ability to characterize locations (see Appendix~\ref{appendix:ablation}). 
This improvement is particularly relevant in settings where historical mobility information is limited, such as when individuals seek novel locations for their activities. 
To benchmark how effectively \method{} leverages contextual cues, we conducted comparative experiments against established mobility flow generation models: Gravity~\citep{zipf_p1_1946}, Deep Gravity~\citep{simini_deep_2021}, and random forest~\citep{cabanas_human_2025}. 
These baselines estimate flows between location pairs using features such as distance, population, and selected facility densities (e.g., shops and transport hubs). 
Flow-based models differ fundamentally from \method{} in both problem formulation and modeling capacity. 
For instance, they cannot capture temporal dynamics or disaggregated mobility. 
To ensure a fair comparison, we focused exclusively on locations that were not observed during training.
The results show that \method{} outperforms the flow generation baselines by a substantial margin across multiple flow metrics (see Appendix~\ref{appendix:flow} and Table~\ref{tab:flow_all}).
We attribute this improved performance in novel locations to the integration of spatial features, including coordinate geometry and detailed POI information, as well as the joint modeling of contextual and behavioral factors. 
While further research is needed to fully understand novelty-seeking mobility, our findings support the hypothesis that comprehensive context modeling is necessary to capture the variability in human behavior.

\subsection{Reproducing mode-specific spatial behavior and co-presence dynamics}\label{sec:application}

Beyond standard validation, we ask whether the generated multi-attribute event sequences support analyses that have so far been difficult or infeasible with existing simulators.
Our simulations generate realistic daily schedules that jointly capture where people go, how they travel, and when activities occur. Earlier tools often rely on predefined rules or cover only part of this behavioral structure, which has limited their use for studying differences across travel modes or patterns of shared presence in everyday settings.
We focus on two representative scenarios with broad relevance to transport planning and urban research: (i) mode-specific spatial visitation, showing how individuals use different travel modes to move through urban space, and (ii) spatio-temporal co-presence, reflecting how individuals co-occur in space and time within shared environments.
Both scenarios are inherently multi-attribute: mode-specific visitation requires coherent joint generation of location and travel mode at the event level, and co-presence requires joint realism in location and time within individual schedules.

\begin{figure*}[!t]
  \centering
  \includegraphics[width=0.99\textwidth]{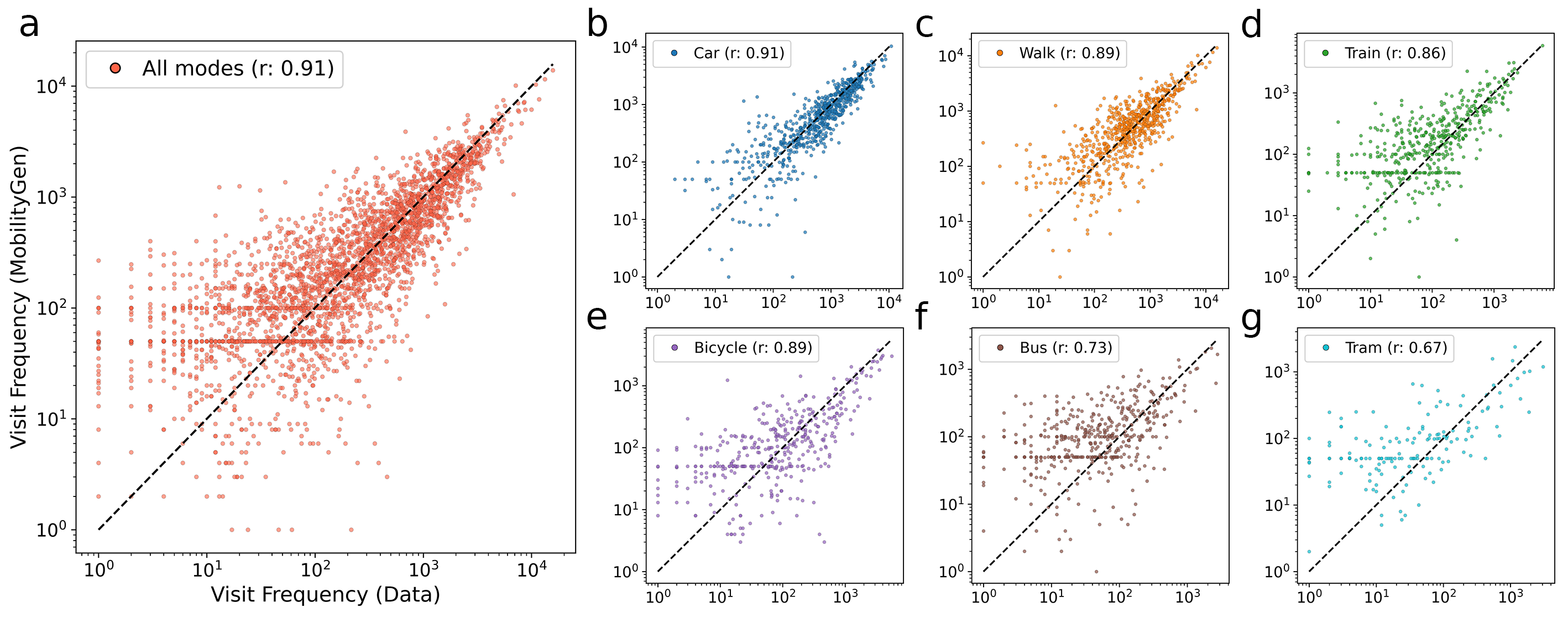}
  \caption{Reproducing location importance across travel modes.
    \textbf{a}. Visit frequencies for grouped locations in real and simulated activity events across all modes, showing strong overall agreement ($r = 0.91$, $P < 10^{-3}$).
    \textbf{b} - \textbf{g}. Mode-specific comparisons for car (\textbf{b}), walk (\textbf{c}), train (\textbf{d}), bicycle (\textbf{e}), bus (\textbf{f}), and tram (\textbf{g}).
    Each dot represents a group of locations with the same visitation frequency observed in the data. Diagonal lines denote perfect agreement. Higher agreement is observed for more frequently used modes (e.g., car and walk), while deviations increase for less common modes (e.g., tram).}
  \label{fig:simulate-importance}
\end{figure*}

Physical spaces serve distinct urban functions and inherently vary in importance. Some locations act as core anchors and attract frequent visits, while others are used more occasionally~\citep{schlapfer_universal_2021, abbiasov_15_2024}. 
\method{} faithfully captures this spatial heterogeneity. 
To demonstrate this, we compare simulated and real activity events by grouping locations according to their visitation frequency. 
The results show strong alignment (Pearson $r=0.91$, $P < 10^{-3}$; Figure~\ref{fig:simulate-importance}a), with important locations consistently retaining their prominence and highlighting the stability of core urban anchors.
Crucially, \method{} enables analysis of how location importance differs across travel modes, a perspective not available in previous approaches.
As shown in Figure~\ref{fig:simulate-importance}b–g, agreement is highest for common modes such as car ($r = 0.91$) and walk ($r = 0.89$), while less frequent modes such as bus ($r = 0.73$) and tram ($r = 0.67$) show greater variation.
These differences reveal how urban space is accessed unevenly across modes and underscore the challenge of simulating less frequent ones. Addressing this gap likely requires targeted modeling strategies and richer contextual data, particularly to better represent sustainable modes that are policy-relevant yet remain underutilized in practice.
By capturing both stable urban anchors and variation in mode-specific access, \method{} enables new analyses of mode-aware demand, supporting applications from infrastructure usage modeling to accessibility assessments and scenario-based policy evaluation.

\begin{figure}[!t]
  \centering
  \includegraphics[width=0.45\textwidth]{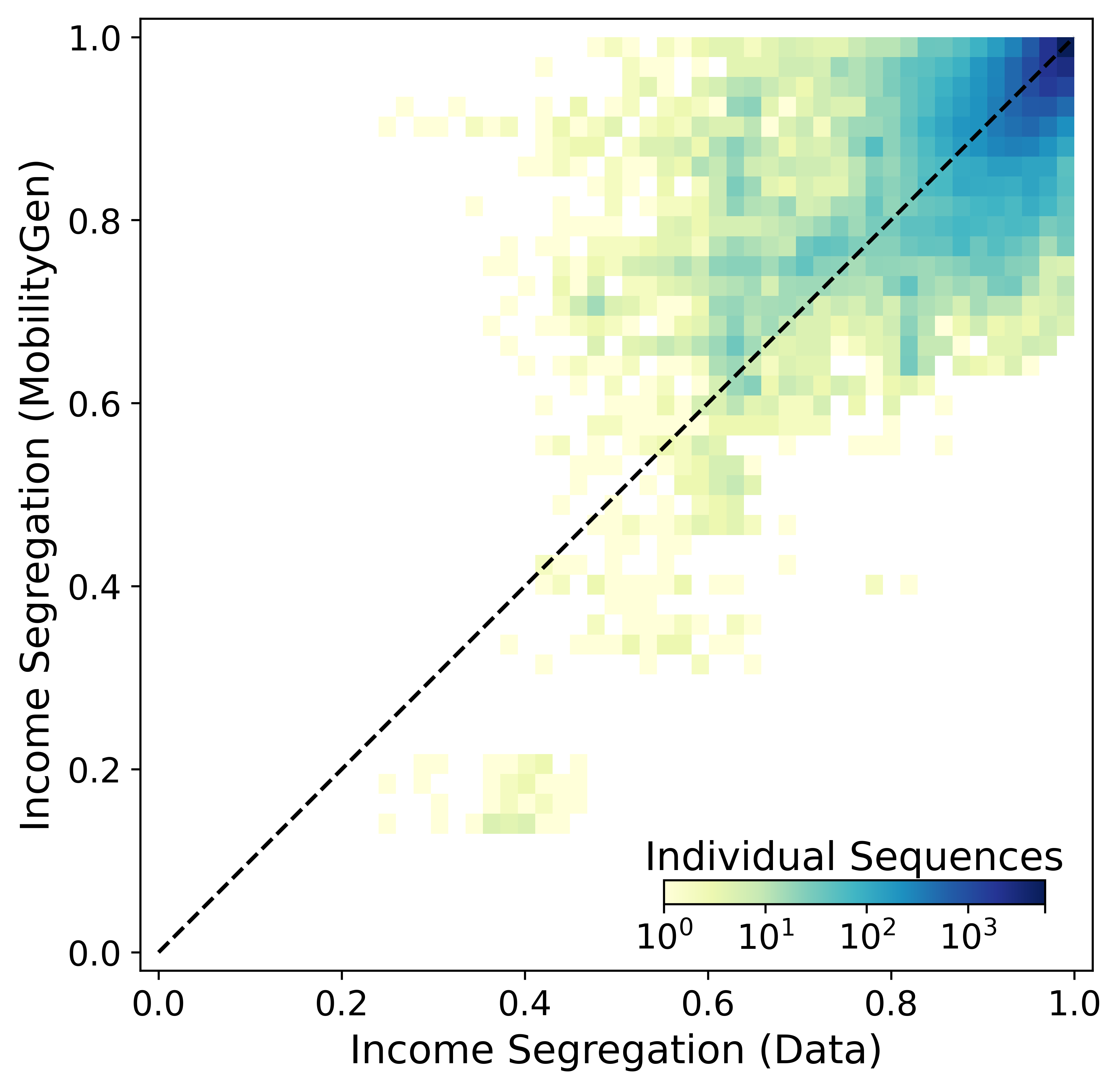}
  \caption{Reproducing individual-level income segregation from co-presence. Heatmap comparing segregation scores computed from observed and simulated activity events. Cell color indicates the number of individuals sharing each observed–simulated score pair. The diagonal line denotes perfect agreement. \method{} shows high agreement with empirical segregation patterns ($r = 0.70$, $P < 10^{-3}$).}
  \label{fig:segregation}
\end{figure}

While spatial visitation captures where individuals go, co-presence patterns reveal whom they encounter, offering complementary insight into the intensity and distribution of interpersonal contact in daily urban life.
A key application of co-presence lies in socio-spatial segregation research, where attention has shifted from static residential indicators to individual-level experienced segregation enabled by fine-grained mobility data~\citep{liao_socio_2025}. 
We assess \method{}’s ability to reproduce these patterns by focusing on income-based segregation. 
We group individuals into four income quantiles and measure, at each location, the time spent by people from different income groups~\citep{moro_mobility_2021} (see Appendix~\ref{appendix:segregation}). 
These measures are then combined into an individual-level segregation score that reflects each person’s exposure to others across income levels.
Simulated and observed segregation scores show strong agreement (Pearson $r = 0.70$, $P < 10^{-3}$; Figure~\ref{fig:segregation}), capturing both overall trends and the variability in co-presence-based income segregation.
A breakdown by income quantiles confirms that \method{} reproduces systematic differences in exposure and substantially outperforms baseline simulators such as TimeGeo and DITRAS, whose separation of space and time limits their ability to capture realistic contact patterns (see Appendix~\ref{appendix:segregation}, Figure~\ref{suppl_fig:segregation}).
This case study illustrates \method{}’s broader value as a generative model of social exposure.
By capturing individual-level interactions in daily schedules, it provides a simulation environment to explore how population-level social mixing evolves under different scenarios. This extends beyond segregation analysis to domains such as information diffusion and infectious disease spread, where spatial co-presence fundamentally shapes outcomes.

\section{Discussion and conclusion}\label{sec:discussion}

Despite growing interest in understanding human movement patterns and their implications, existing simulation models fail to capture the full spatio-temporal variability of mobility behavior.
Here, we address this gap by adapting diffusion-based generative modeling to activity-travel event sequences, generating multi-attribute mobility events over days and weeks.
We instantiate this approach in \method{}, which provides strong alignment with empirical patterns across key mobility indicators and captures coupled structure across time, space, and mode that earlier approaches struggle to represent.
Its generative capability is reflected in coherent held-out activity-travel sequences, plausible visits to previously unobserved locations, and emergent individual- and population-level patterns.
Because the model produces complete multi-attribute trajectories, it supports downstream analyses such as how urban space is accessed differently across travel modes and how co-presence dynamics shape social exposure, including income-based segregation.
Taken together, these capabilities show how deep generative modeling can extend mobility simulation beyond aggregate-statistic alignment to fine-grained analyses of mobility, accessibility, and social exposure.

Our data-driven framework extends event-level representations into a generative modeling setting.
By integrating behavioral attributes into a shared latent space, \method{} captures empirical dependencies across multi-day sequences that are difficult to specify exhaustively through predefined rules.
This positions \method{} as a complement to behaviorally grounded activity and transport demand models: where those models encode relationships a priori, \method{} learns them from observed sequences.
These learned representations are also inspectable, with the embedding space analysis revealing behaviorally meaningful structure as latent geometry, a possible substrate for further behavioral investigation.
Looking ahead, \method{} could be paired with structural behavioral models in hybrid approaches that combine learned representations with explicit behavioral assumptions, supporting applications such as household activity participation and targeted mobility interventions.
Its flexibility also supports richer contextual information, from socio-demographic characteristics to social networks and environmental factors such as weather.
The held-out-individual analysis delineates a boundary of the current framework: activity-travel regularities transfer more robustly to individuals absent from training, whereas spatial generation remains sensitive to source-location coverage. 
Extending the framework toward settings with limited or no observed individual history is therefore a practically important direction, since detailed mobility data are typically available only for small subsets of a population, whereas many applications require mobility estimates at population scale. 
Together, these extensions position \method{} as a bridge between mobility research and disciplines such as social, health, and environmental sciences, a step toward an integrative framework for interdisciplinary mobility modeling.

From a practical standpoint, \method{} generates long-term, realistic, and multifaceted activity schedules for individuals using GNSS tracking data, reducing the need for labor-intensive traditional travel surveys.
Generative mobility models offer several distinct advantages: they enable the creation of synthetic datasets at scales far beyond those of real-world collections and mitigate privacy concerns when analyzing or sharing disaggregated data.
These large, openly accessible datasets, in turn, could promote the standardization and intercomparison of movement analysis methods.
Finally, access to fine-grained synthetic mobility data could unlock previously unattainable simulation scenarios and studies in support of sustainable transport, urban planning, public health, and beyond.


\section*{Data availability}

Raw data for the MOBIS dataset are not publicly available due to privacy considerations, but aggregated trip-level data are available to reviewers upon request.
OpenStreetMap POIs used for context were downloaded from Geofabrik (region: Europe/Switzerland; snapshot: late 2022) and can be downloaded from \url{https://download.geofabrik.de/}.

\section*{Code availability}

All code to preprocess movement data and train \method{} is available at \url{https://github.com/mie-lab/mobility_generation} (Apache-2.0 license). Analyses and visualizations were performed in Python and QGIS 3.22.

\section*{Declaration of generative AI and AI-assisted technologies in the writing process}

During the preparation of this work, the author(s) used GPT-5 in order to improve the language, grammar, and readability. After using this tool/service, the author(s) reviewed and edited the content as needed and take(s) full responsibility for the content of the publication.


\newpage
\beginsupplement
\section*{Appendix}

\setcounter{section}{0}
\section{Simulating individual preferences in location visits}
\subsection{Comparison with microscopic mobility models}\label{appendix:loc}

\begin{figure}[!htb]
  \centering
  \includegraphics[width=0.99\textwidth]{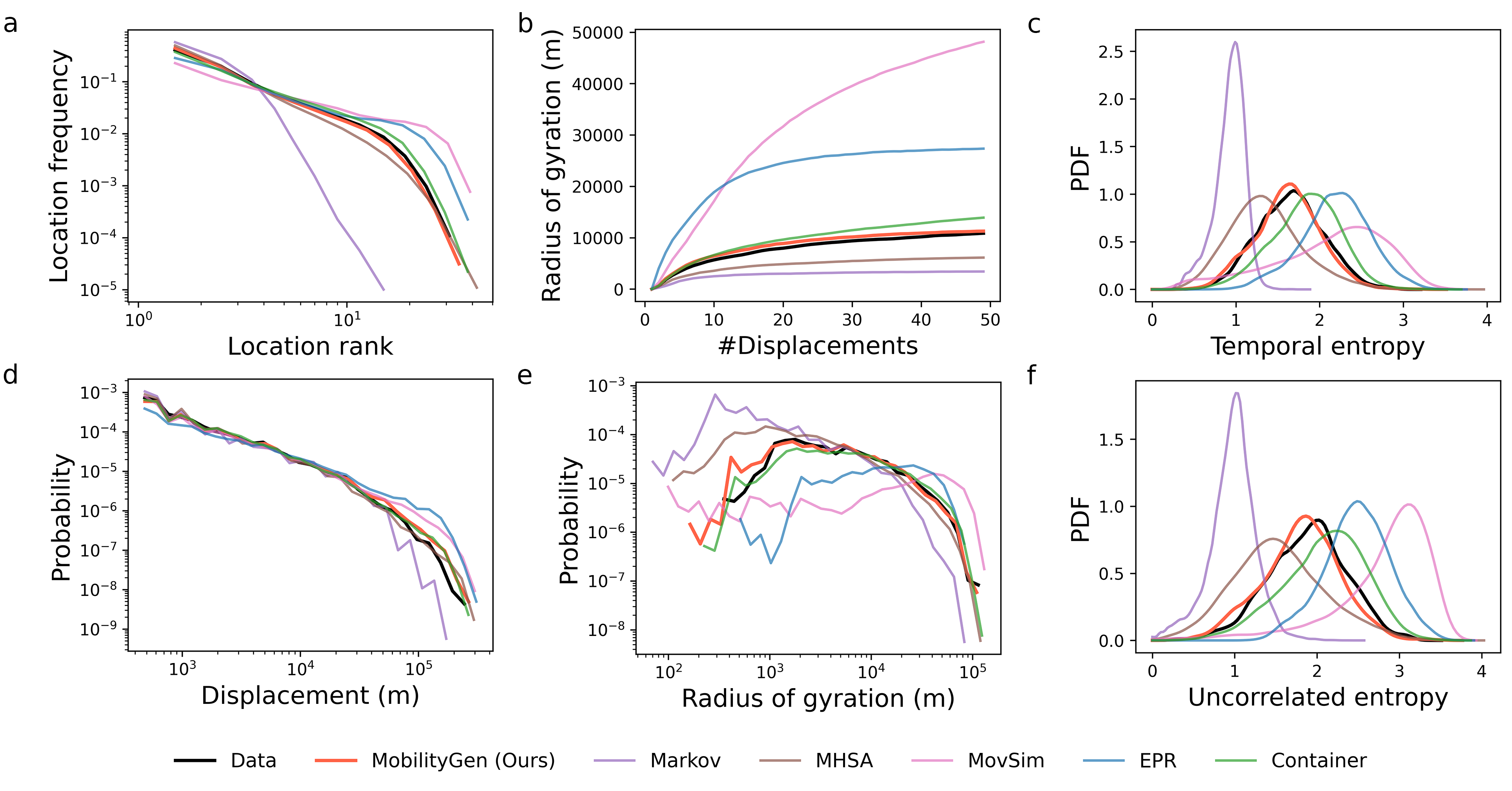}
  \caption{Evaluating microscopic mobility models with location metrics.
  \textbf{a}. Rank-frequency distribution of visited locations.
  \textbf{b}. Median radius of gyration as a function of the number of displacements.
  \textbf{c}. Distribution of temporal mobility entropy across individuals.
  \textbf{d}. Distribution of displacements for the entire population.
  \textbf{e}. Distribution of radius of gyration across individuals.
  \textbf{f}. Distribution of uncorrelated mobility entropy across individuals.
  These metrics are compared between real data (black) and traces generated by \method{} (red), Markov (purple), MHSA (brown), MoveSim (pink), EPR (blue), and Container (green).}
  \label{suppl_fig:loc_metric_all}
\end{figure}

To evaluate the ability of \method{} to replicate fundamental properties of individual location behavior, we compare it with representative microscopic mobility models spanning mechanistic, predictive, and generative approaches. These include: (i) classical mechanistic models such as the exploration and preferential return (EPR) model~\citep{song_modelling_2010} and the Container model~\citep{alessandretti_scales_2020}; (ii) sequence prediction models, including a first-order Markov chain~\citep{gambs_next_2012} and a multi-head self-attention (MHSA)-based neural network~\citep{Hong_context_2023}; and (iii) the deep generative MoveSim model~\citep{feng_learning_2020}. The design and implementation details of each model are described in the following sections.

\begin{itemize}
    \item The EPR model is a well-established microscopic mobility model designed to reproduce scaling laws in the evolution of distinct locations and their visitation frequencies over time. We estimate model parameters from the training data. For each test sequence, we determine the number of unique locations \( S \) visited and the corresponding visit counts \( f_i \) for each location \( i \), and then simulate 50 displacement steps. At each step, the individual either explores a new location with probability \( p^{\text{new}} = \rho S^{-\gamma} \), where $\rho$ and $\gamma$ are parameters, or returns to a previously visited location with the complementary probability \( 1 - p^{\text{new}} \). In exploration steps, a new location is determined by sampling a displacement \( \Delta r \) from the empirically estimated jump length distribution \( P(\Delta r) \). For return steps, the probability of selecting location \( j \), denoted as \( \Pi_j \), is proportional to its past visitation frequency, i.e., \( \Pi_j \propto f_j \).
    \item The Container model describes how individuals allocate time and move across nested spatial scales \citep{alessandretti_scales_2020}. The model is built on a hierarchical structure $H$ with $L$ levels, where each level groups locations into increasingly broader spatial containers. Each container is associated with an attractiveness score $a$, and mobility decisions are governed by an $L \times L$ displacement matrix $D$. The probability of moving from the current location $j$ to a destination $k$ is given by: 
    \begin{equation}
        \label{equation:container}
        p_{D_{j,h}, D_{j,k}}\prod_{l \le D_{j,k}}a(k_l)
    \end{equation}
    Here, the first term $p_{D_{j,h}, D_{j,k}}$ denotes the probability of traveling a distance $D_{j,k}$, given that the individual is currently at distance $D_{j,h}$ from their home location $h$. The second term captures the probability of choosing location $k$ at that level distance, where $a(k_l)$ represents the attractiveness of a container at level $l$ that contains location $k$. Parameters $H$, $a$, and $D$ are estimated for each individual via maximum likelihood estimation on the training data~\citep{alessandretti_scales_2020}. We generate 50 displacements per individual for evaluation on the test set.  
    
    \item Markov model. Classical models for location prediction represent individual mobility as a Markov chain~\citep{AshbrookS02}, where locations are treated as states and transitions between them are encoded in a Markov transition matrix. We implement a first-order Mobility Markov Chain (1-MMC)~\citep{gambs_next_2012}, in which the probability of visiting the next location depends only on the current one. During generation, the next location is sampled from the top three most likely locations based on the learned transition probabilities. This process is repeated until the generated sequence reaches the target length of 50 steps.
    
    \item MHSA-based model. Recent advances in next location prediction leverage deep sequential models that represent locations as latent embeddings and learn transition dynamics via attention-based architectures~\citep{luca_survey_2021}. We train an MHSA network to autoregressively generate location sequences. At each step, the next location is sampled using a combination of top-$k$ and nucleus sampling~\citep{holtzman_curious_2020}. The top-$k$ strategy restricts sampling to the $k$ most probable candidates, while nucleus sampling limits it to the smallest set of locations whose cumulative probability exceeds a predefined threshold $p$. We set $k = 200$ and $p = 0.99$. The selected location is appended to the sequence and used as input for the next prediction step. This procedure is repeated for 50 steps.
    
    \item MoveSim. As a baseline for generative modeling of individual location choices, we implement MoveSim, a sequence-to-sequence model based on generative adversarial networks (GANs). The model consists of a generator that simulates location sequences using mobility priors, and a discriminator that distinguishes between real and generated sequences. These networks are trained jointly in an adversarial framework using a min-max objective. Unlike the original implementation, we exclude auxiliary features such as historical transitions, physical distances, and functional similarity, as they require pairwise location matrices that are computationally infeasible to construct given the large number of unique locations in our dataset.
\end{itemize}

\addtolength{\tabcolsep}{-1pt}
\begin{table}[!b]
    \caption{Model performance in replicating location visit distributions. The log-likelihood ratio measures the likelihood that the real data was generated by each model relative to \method{}, with positive values favoring \method{} (all \( P < 10^{-3} \), except for $P(f_{\Delta r})$ in MHSA, where $P=0.66$). The Wasserstein distance assesses the dissimilarity between generated and real location metric distributions, with higher values indicating greater divergence. Results are shown for number of visits per location (\( f_k \)), jump length (\( f_{\Delta r} \)), radius of gyration (\( f_{r_{u}} \)), uncorrelated entropy (\( S_{\text{unc}} \)), and temporal entropy (\( S_{\text{temp}} \)); best-performing values are shown in bold.}
    \label{tab:distance_loc}
    \centering
    \begin{tabular}{@{}lcccccccccc@{}}
    \toprule 
              & \multicolumn{5}{c}{Log-likelihood} & \multicolumn{5}{c}{Wasserstein distance} \\ \cmidrule(lr){2-6} \cmidrule(lr){7-11}
              &  $P(f_k)$       & \( P(f_{\Delta r}) \) &  \( P(f_{r_{u}}) \) & \( P(S_{\text{unc}}) \) & $P(S_{\text{temp}})$ &  $f_k$  &  \( f_{\Delta r} \) &  \( f_{r_{u}} \) & \( S_{\text{unc}} \)            & $S_{\text{temp}}$      \\\midrule
    EPR       &  114195         & 166326 & 31364  & 74711   & 30298   & 2.62 & 15882 & 12825 & 0.60 & 0.54   \\
    Container &  12504          & -6568 & 946     & 2741    & 3549    & 0.68 & \textbf{1146} & 3578 & 0.20 & 0.18  \\
    Markov    &  796846         & 2697 & 16530    & 329477  & 2728649 & 2.03 & 2223 & 10102 & 0.91 & 0.73     \\
    MHSA      &  66983          & -115 & 4352     & 8355    & 9616    & 0.85 & 1491 & 5137  & 0.35 & 0.30  \\
    MoveSim    &  268850         & 35384 & 79922   & 48838   & 22837   & 5.04 & 8128 & 34096  & 0.92& 0.51   \\
    \method{} (Ours) &  -       & - & -           & -       &  -   & \textbf{0.26} & 2318 & \textbf{575} & \textbf{0.08} & \textbf{0.04}   \\\bottomrule
    \end{tabular}
    
\end{table}
\addtolength{\tabcolsep}{1pt}

We evaluate the performance of all models using established location-based metrics. In addition to those reported in the main text, we include the jump length distribution~\citep{brockmann_scaling_2006} and the uncorrelated entropy~\citep{song_limits_2010}. Results are shown in Figure~\ref{suppl_fig:loc_metric_all}. 
To assess model fit, we compute the log-likelihood ratio, comparing the likelihood of the observed data under each model relative to \method{}. We additionally calculate the Wasserstein distance to quantify discrepancies between empirical and model-generated distributions for each metric (Table~\ref{tab:distance_loc}). Overall, \method{} provides the closest approximation to the empirical data across all evaluated metrics, except for the jump length distribution, where it tends to overestimate the frequency of longer displacements (Figure~\ref{suppl_fig:loc_metric_all}d). 
Moreover, the median evolution of the radius of gyration across individuals, \( r_u(k) \), over displacement steps \( k \), is well described by logarithmic growth (Figure~\ref{suppl_fig:loc_metric_all}b). The fit \( r_u(k) \propto \alpha \log(k) \), with parameters \( \alpha \) derived from \method{} ($\alpha_{\text{\method}} = 2967 \pm 29$), aligns most closely with the real data ($\alpha_{\text{true}} = 2875 \pm 35$), while the baselines exhibit notably different values ($\alpha_{\text{epr}} = 6431 \pm 185$, $\alpha_{\text{container}} = 3784 \pm 68$,
$\alpha_{\text{markov}} = 796 \pm 26$, $\alpha_{\text{mhsa}} = 1541 \pm 12$,
$\alpha_{\text{moveSim}} = 14976 \pm 462$).

\begin{figure}[!b]
  \centering
  \includegraphics[width=0.99\textwidth]{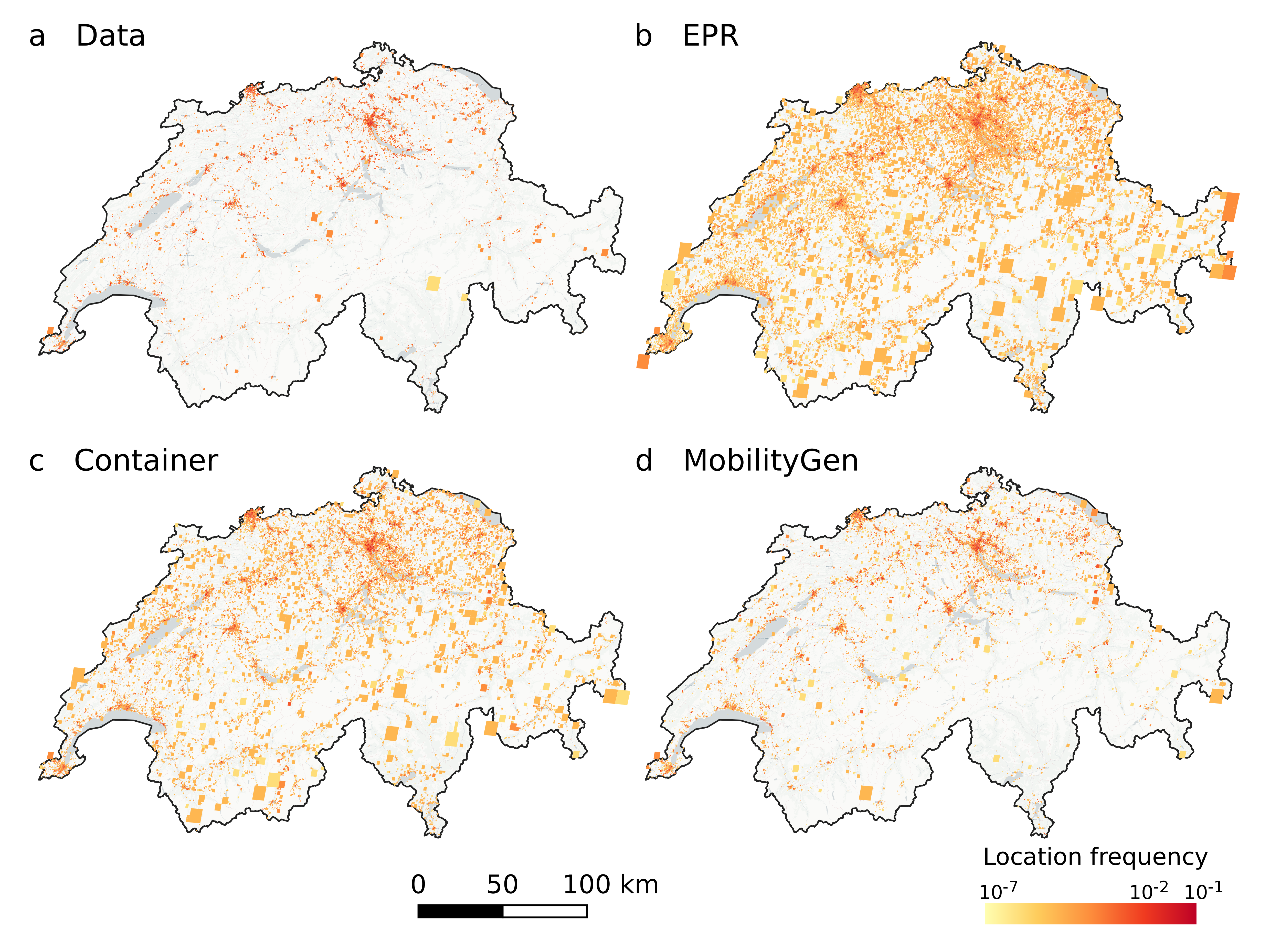}
  \caption{Comparison of spatial visitation patterns across models. Aggregated visit frequencies from real data (\textbf{a}) are compared with those generated by three baseline mobility models: EPR (\textbf{b}), Container (\textbf{c}), and \method{} (\textbf{d}). While all models recover broad national-scale trends, \method{} more accurately reproduces both high-density urban visitation and spatial dispersion into surrounding regions.}
  
  \label{suppl_fig:compare_full}
\end{figure}

\begin{figure}[!htb]
  \centering
  \includegraphics[width=0.99\textwidth]{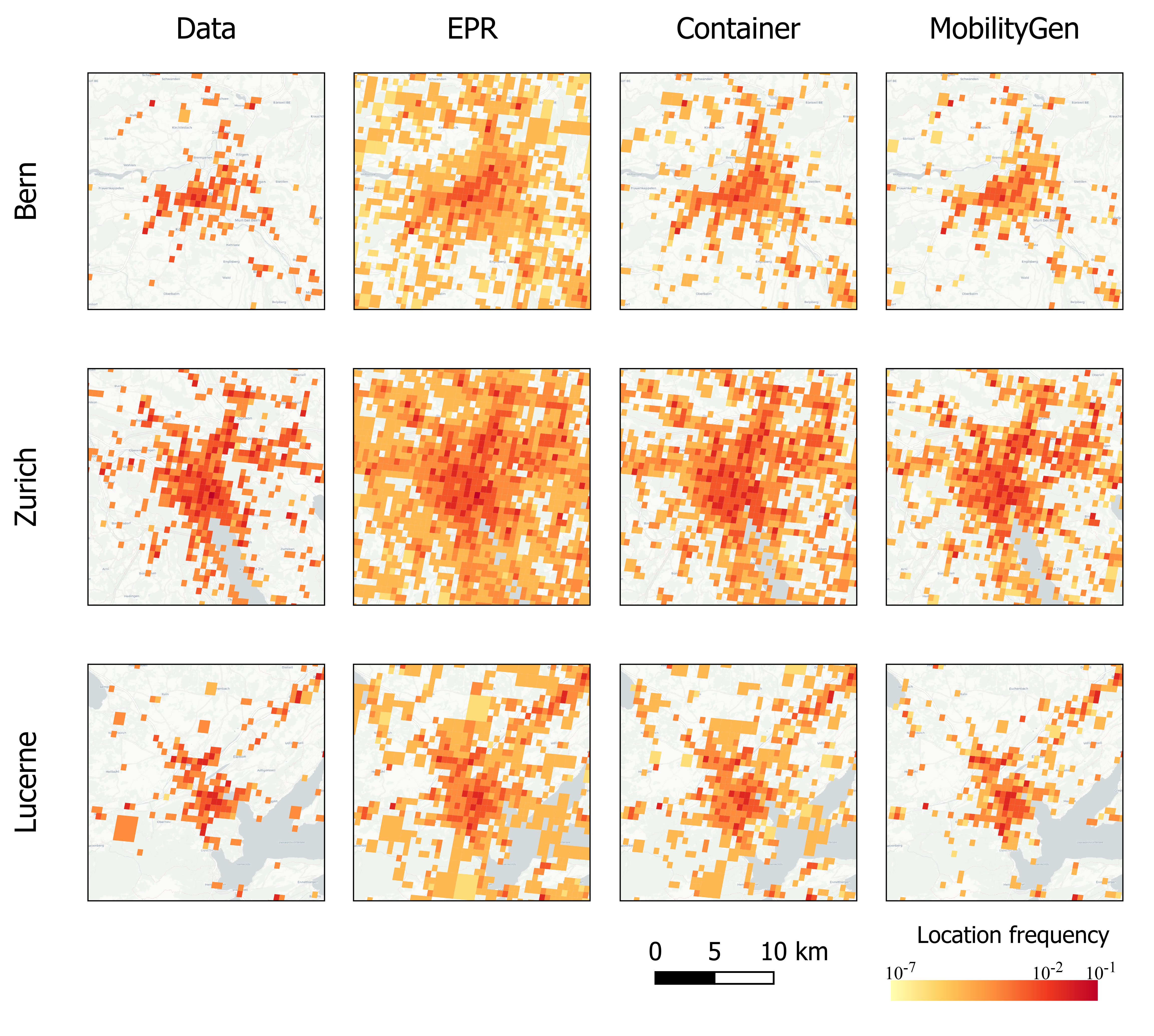}
  \caption{Urban-level comparison of spatial visitation patterns across models. Real and simulated visit frequencies are compared across Bern (top row), Zurich (middle row), and Lucerne (bottom row). Consistent with national-level patterns, \method{} best captures fine-grained spatial visitation within dense urban cores and across peripheral areas.}
  
  \label{suppl_fig:compare_detail}  
\end{figure}

Beyond matching marginal distributions, we also examine whether each model reproduces the geographic concentration and dispersion of visits at population scale. Figure~\ref{suppl_fig:compare_full} compares aggregated visitation intensity across Switzerland for the empirical data and each microscopic model, while Figure~\ref{suppl_fig:compare_detail} provides the corresponding urban-scale comparisons for Bern, Zurich, and Lucerne. To summarize these maps quantitatively, we report the entropy of the visit-frequency distribution across locations (Table~\ref{tab:entropy_loc}), where higher values indicate broader spatial dispersion. The baselines either concentrate visits too strongly or disperse them too widely relative to the data, whereas \method{} yields dispersion values that are closer to the empirical patterns at both national and city scales, while still showing a tendency toward slightly higher dispersion in several cases.

\begin{table}[!htb]
    \caption{Spatial dispersion in visitation patterns. Entropy values are reported for the real data and for location traces generated by each microscopic mobility model. Higher entropy indicates broader spatial dispersion of visits. Values below those of the real data suggest insufficient reproduction of exploratory behavior, while values above indicate misalignment with population-level location preferences.}
    \label{tab:entropy_loc}
    \centering
    \begin{tabular}{@{}lcccc@{}}
    \toprule 
                    &  National  & Bern &  Zurich              & Lucerne      \\\midrule
    EPR             &  1.20        & 3.89   & 5.04 & 3.50   \\
    Container       &  0.91        & 3.13   & 4.60 & 3.18  \\
    Markov          &  0.17        & 1.03   & 2.07 & 1.34     \\
    MHSA            &  0.50        & 2.79   & 4.05 & 2.62  \\
    MoveSim          &  0.67        & 3.21   & 4.30 & 2.58   \\
    \method{} (Ours)&  0.53        & 2.49   & 3.99 & 2.71   \\\midrule
    Data            &  0.33        & 1.89   & 3.37 & 2.11   \\\bottomrule
    \end{tabular}
    
\end{table}

Collectively, the results show that \method{} more accurately captures key dimensions of location behavior, including visitation frequency, travel distance, and spatio-temporal spread, than traditional microscopic models.

\subsection{Quantifying similarity and diversity in individual location sequences}\label{appendix:language}

To complement classical mobility metrics, we evaluate the simulated location sequences using metrics adapted from language modeling, which quantify both their similarity to and diversity from the real sequences. 
Similarity is measured using the Bilingual Evaluation Understudy (BLEU) score~\citep{Papineni_bleu_2002}, an n-gram-based metric that compares each generated sequence to its corresponding ground truth, with higher values indicating greater similarity. We implement the smoothed sentence-level version of BLEU at n-gram levels 1 through 4 (BLEU-1 to BLEU-4), commonly used in sequence generation tasks.
To evaluate intra-sequence diversity, we use the distinct unigram (dist-1) metric~\citep{gong_diffuseq_2023}, which measures the proportion of unique locations within a sequence. Lower dist-1 values indicate higher repetition and less diversity.
At the sequence level, we compute the diverse 4-gram (div-4) metric~\citep{Deshpande_fast_2019}, which captures the proportion of distinct 4-grams in a sequence. Higher div-4 values indicate richer generation diversity.
Evaluation results for all implemented microscopic mobility models are presented in Table~\ref{tab:variability}.
Overall, \method{} produces location visit sequences that closely mirror real-world patterns, while avoiding input replication and preserving a realistic level of behavioral diversity.

While these metrics are informative, they should be interpreted with caution:
(i) In contrast to language modeling, where multiple semantically valid reference sentences may exist, mobility datasets typically provide only one observed realization of an individual’s behavior, despite the existence of many plausible alternatives.
(ii) Realistic mobility behavior involves a balance between exploration and habitual returns. As such, high diversity is not inherently desirable, and diversity metrics alone may not fully capture the plausibility of generated sequences.

\begin{table}[!htb]
    \caption{Sequence similarity and diversity with language modeling metrics. The BLEU score quantifies similarity between generated and real location sequences, with higher values indicating better alignment. Distinct unigrams (dist-1) and diverse 4-grams (div-4) measure sequence diversity at the intra-sequence and inter-sequence levels, respectively, where higher values indicate greater variability.}
    \label{tab:variability}
    \centering
    \begin{tabular}{@{}lcccc@{}}
    \toprule 
                     &  BLEU       & dist-1 &  div-4              \\\midrule
    Markov           &  0.28      & 0.08  & 0.02 \\
    MHSA             &  0.24      & 0.21  & 0.29 \\
    MoveSim           &  0.002      & 0.51  & 0.60 \\
    EPR              &  0.09      & 0.42  & 0.77 \\
    Container        &  0.19      & 0.31  & 0.57 \\
    \method{} (Ours) &  0.26      & 0.23  & 0.34 \\\bottomrule
    \end{tabular}
    
\end{table}

\section{Generating realistic activity event sequences}
\subsection{Comparison with spatio-temporal mobility simulators}\label{appendix:attribute}

\begin{figure}[!htb]
  \centering
  \includegraphics[width=0.99\textwidth]{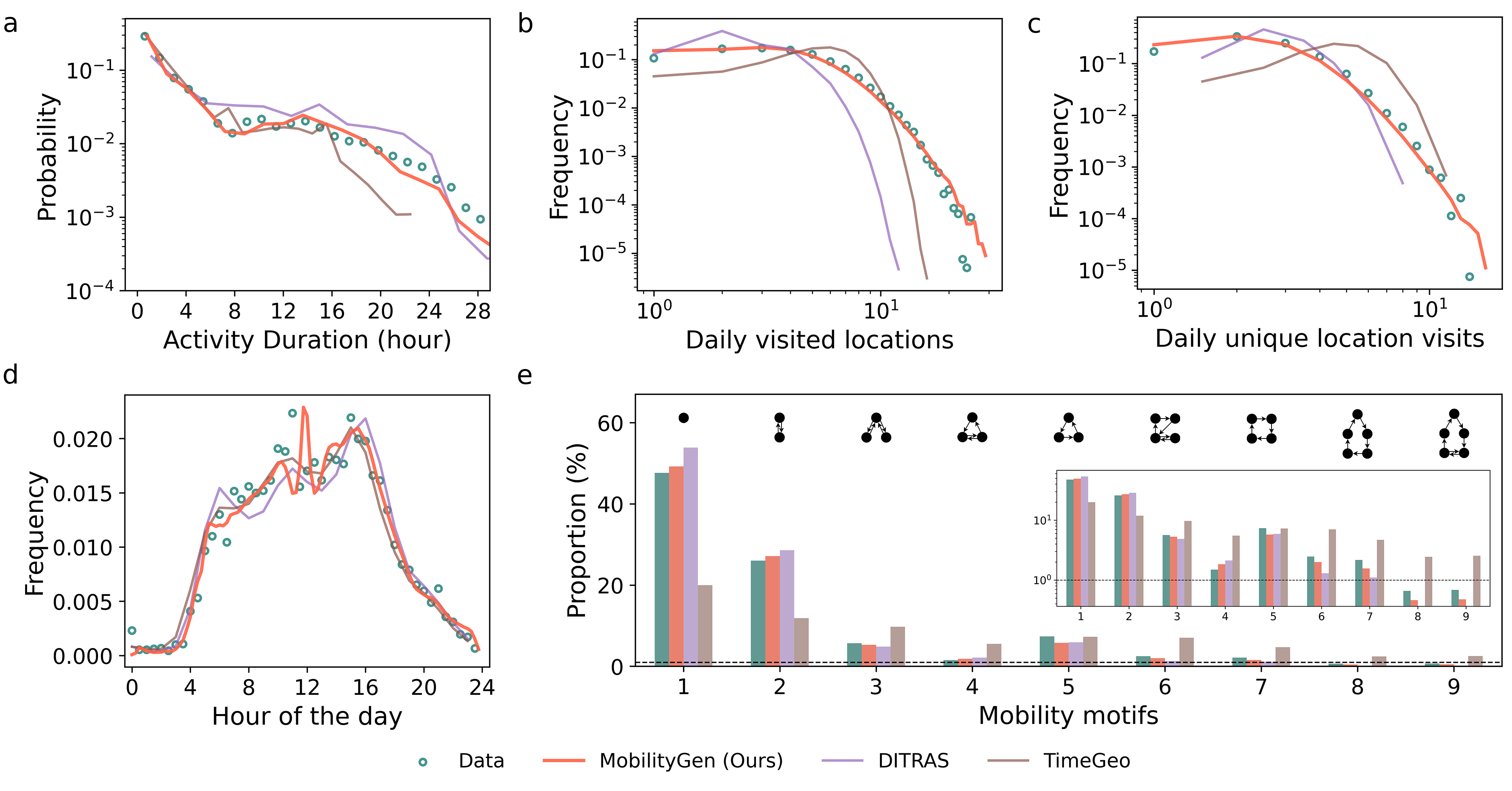}
  \caption{Generated events replicate individual activity-travel patterns and interdependencies.
  \textbf{a}. Distribution of activity durations.
  \textbf{b}. Distribution of daily visited locations.
  \textbf{c}. Distribution of daily unique visited locations.
  \textbf{d}. Distribution of activity start times throughout the day.
  \textbf{e}. Distribution of mobility motifs, with motif diagrams shown above. The dashed line marks the 1\% threshold as a visual guide, and the inset presents the same distribution on a log scale.
  Metrics are compared between real data (green) and activity events generated by \method{} (red), DITRAS (purple), and TimeGeo (brown).}
  \label{suppl_fig:attribute_metric_all}
\end{figure}

To evaluate the effectiveness of \method{} in reproducing spatio-temporal activity-travel patterns, we benchmark its performance against two state-of-the-art mobility frameworks: the Diary-based Trajectory Simulator (DITRAS)~\citep{pappalardo_data_2018} and TimeGeo~\citep{jiang_timegeo_2016}. Detailed descriptions of their methodologies and implementations are provided below.
\begin{itemize}
    \item DITRAS decouples temporal pattern generation from spatial location assignment by modeling temporal dynamics with a mobility diary generator and assigning these diaries to physical locations using the density-EPR model~\citep{pappalardo_returners_2015}. The diary generator employs a time-aware Markov transition matrix to estimate transition probabilities between ``typical'' locations (e.g., home) and ``non-typical'' locations, along with the corresponding durations of stay. Non-typical locations are then assigned using the density-EPR mechanism, an extension of the classic EPR model that incorporates a relevance factor during exploration. We used the DITRAS implementation from scikit-mobility to align with our data format. Based on the training dataset, we estimated the population-level Markov transition matrix and calibrated individual-specific density-EPR parameters. Location relevance was determined by empirical visit frequencies. We adopted a temporal resolution of one hour for simulation.
    
    \item TimeGeo. Originally developed to reconstruct high-resolution spatio-temporal mobility trajectories from sparse digital traces, TimeGeo incorporates three individual-specific parameters to characterize temporal movement dynamics: the weekly number of home-based tours, the dwell rate, and the burst rate. These parameters govern time-dependent travel tendencies through a Markov decision process that distinguishes movement patterns based on an individual's location state. Spatial decisions are modeled using a rank-based EPR mechanism, which selects exploration locations based on their proximity rank relative to the current location, rather than absolute geographic distance. We estimated temporal and spatial parameters for each individual using the training dataset. For each test sequence, the most frequently visited location was designated as the home location, and the model was simulated at a temporal resolution of one hour.

\end{itemize}

\begin{table}[!b]
    \caption{Alignment of behavioral patterns and interdependencies. The Wasserstein distance quantifies dissimilarities between metric distributions derived from simulated and real activity events, with higher values indicating greater divergence. Source sequence results are included as a reference baseline. Results are reported for activity duration ($f_d$), activity start times ($f_{t}$), daily visited locations ($f_k^{\text{day}}$), unique daily visited locations ($f_{\left| k \right|}^{\text{day}}$), mobility motifs ($f_{\text{motifs}}$), and travel mode ($f_{m}$); best-performing values are shown in bold.}
    \label{tab:distance_all}
    \centering
    \begin{tabular}{@{}lcccccc@{}}
    \toprule 
                        & $f_d$   & $f_{t}$  & $f_k^{\text{day}}$ & $f_{\left| k \right|}^{\text{day}}$ & $f_{\text{motifs}}$ & $f_{m}$ \\\midrule
    DITRAS              &  2.48   & 1.56  & 1.51  &  0.39 & 0.30 & --          \\
    TimeGeo             &  1.92   & 1.33  & 1.33  &  1.33 & 1.37 & --          \\ 
    \method{} (Ours)    &  \textbf{0.47}  & \textbf{0.52}    & \textbf{0.29} &  \textbf{0.22} & \textbf{0.14} & 0.048          \\ \midrule
    Source sequence     &  0.58   & 1.11 & 0.66 & 0.29 & 0.27 & \textbf{0.043}          \\ \bottomrule
    \end{tabular}
    
\end{table}

We assess the performance of these models using mobility metrics that characterize activity-travel patterns. 
To align the hourly outputs of DITRAS and TimeGeo with \method{}’s event-sequence formulation, we converted their simulated hourly state sequences into activity event sequences by merging consecutive records assigned to the same location. 
We then retained the first 50 generated events from each model and compared them with the first 50 real future events, ensuring that all models are evaluated over the same event horizon.
In addition to the metrics reported in the main text, we introduce two supplementary measures: the number of unique location visits per day and the distribution of activity start times. Evaluation results are presented in Figure~\ref{suppl_fig:attribute_metric_all}. To quantify model performance, we compute the Wasserstein distance between the distributions of each metric derived from the generated activity events and those observed in the real data (Table~\ref{tab:distance_all}). Metrics calculated from the source sequence are included as a strong reference baseline. However, it is important to account for variability in human activity behavior, as source sequences differ in both the duration and timing of activity events.

\method{} consistently outperforms all other models in capturing individual activity-travel patterns and interactions, demonstrating a substantial margin of improvement. Moreover, it simulates event sequences that are as accurate as, and in some cases even more accurate than, those derived from the source sequences.

\subsection{Analysis of learned attribute embeddings}\label{appendix:embedding}

\begin{figure}[!h]
  \centering
  \includegraphics[width=0.99\textwidth]{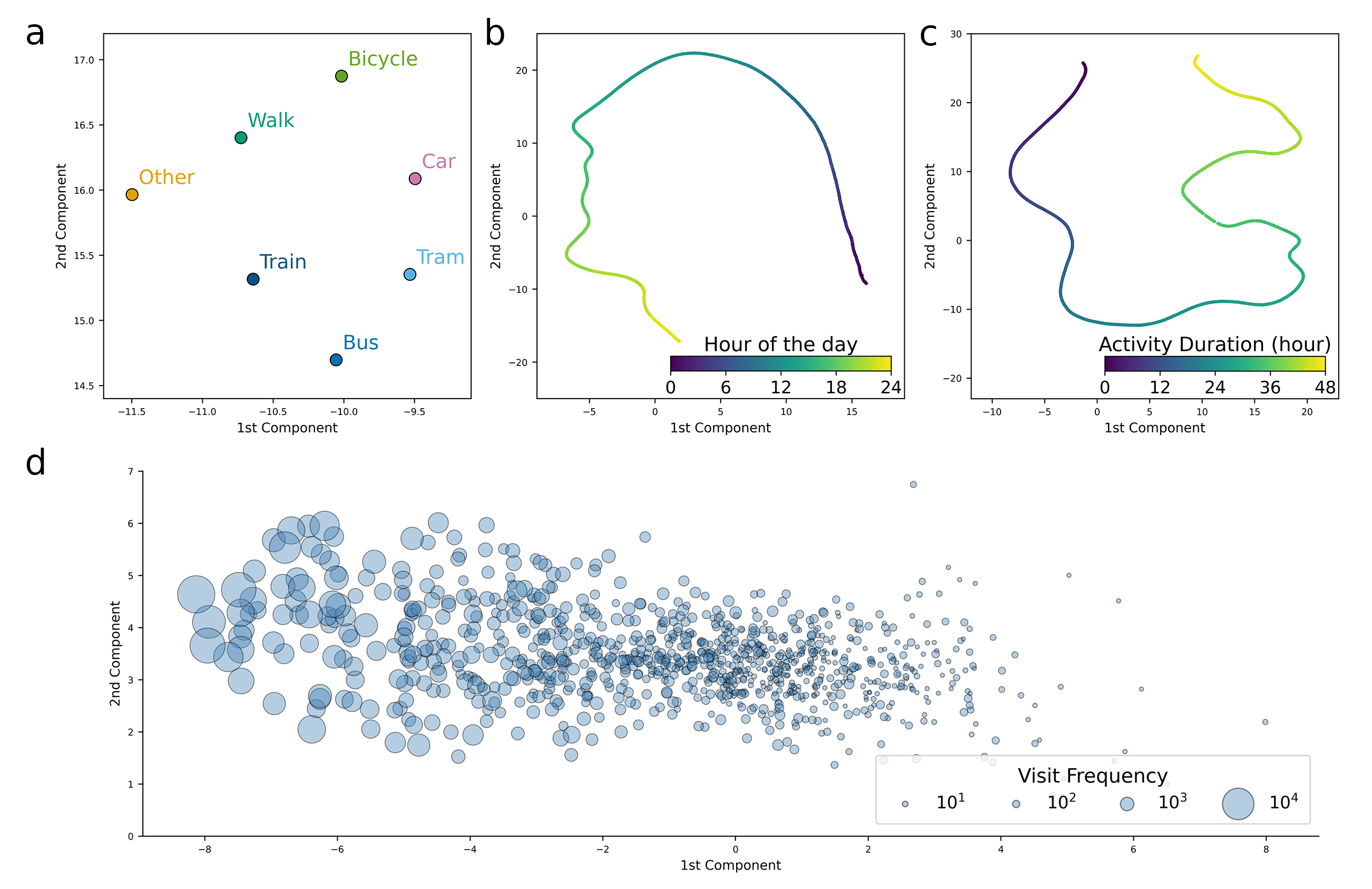}
  \caption{Visualization of learned activity embeddings. 2D projections of the 128D attribute embeddings, obtained using the densMAP algorithm (min\_dist=0.1, dens\_lambda=2.0).
  \textbf{a}. Travel mode embeddings reveal interpretable structure: similar modes (walk and bike; bus, tram, and train) are positioned closely, while ``other'' appears isolated.
  \textbf{b}. Time embeddings form a continuous trajectory that progresses over the 24-hour day, with neighboring hours mapped to adjacent regions.
  \textbf{c}. Duration embeddings follow a similarly ordered curve, reflecting the gradual increase from 0 to 48 hours.
  \textbf{d}. Location embeddings are structured by visit frequency, with marker size denoting log-scaled frequency. Higher-frequency locations cluster toward one end of the space.}
  \label{suppl_fig:embeddings}
\end{figure}

\begin{figure}[!b]
  \centering
  \includegraphics[width=0.99\textwidth]{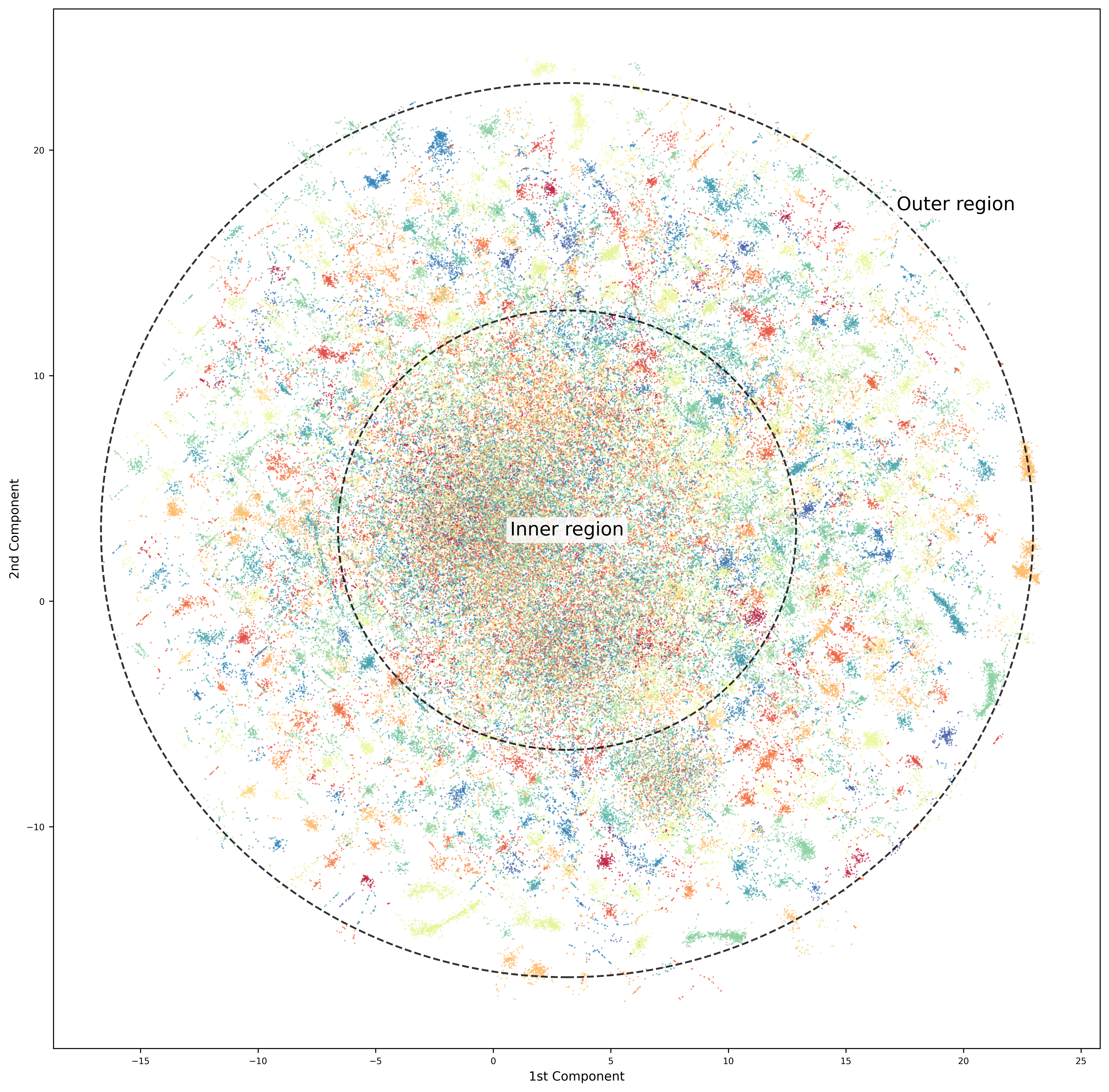}
  \caption{Visualization of learned event embeddings. Each point represents a mobility event defined by a unique combination of travel mode, time, duration, and location, with colors indicating the visited location. The embedding space shows a clear structural distinction between two loosely defined regions: an outer region where events form location-based clusters, and an inner region where events are mixed without clear spatial separation. The circular boundaries serve as visual guides to illustrate the transition from disordered to clustered patterns. This contrast suggests that the model distinguishes between different types of mobility behavior.}
  \label{suppl_fig:event_embedding}
\end{figure}

We examine the structure of the learned attribute embeddings to uncover how \method{} internalizes and organizes behavioral attributes. 
Each 128-dimensional embedding is projected onto a two-dimensional space using the densMAP method, as implemented in the Python library UMAP~\citep{mcinnes2018umap_software}. 
We use the default parameters (\(min\_dist=0.1\), \(dens\_lambda=2.0\)) and confirm that alternative parameter combinations yield qualitatively consistent embedding structures. 
Two-dimensional visualizations are generated for both single attributes and composite activity events, computed according to Eq.~\ref{equation:attribute_embed}.

Visualizations for travel mode, time, duration, and location are shown in Fig~\ref{suppl_fig:embeddings}. 
Travel modes with similar characteristics are positioned in close proximity within the embedding space; for example, walk and bike appear nearby, as do train, bus, and tram. 
In contrast, the ``other'' category appears isolated, reflecting its dissimilarity from all other modes. 
The time and duration embeddings in Figure~\ref{suppl_fig:embeddings}b and c form ordered, coherent trajectories, revealing that \method{} captures the continuous structure inherent in these attributes. 
For locations in Figure~\ref{suppl_fig:embeddings}d, the model organizes places according to their empirical visit frequencies in the training set, even though this frequency information is not explicitly observable from individual event sequences. 
This suggests that the model encodes the relative importance of locations based on usage patterns and implicitly captures the characteristic rank-frequency distribution observed in human mobility.

We next examine composite event embeddings (Figure~\ref{suppl_fig:event_embedding}). Each point represents a mobility event defined by a combination of travel mode, time, duration, and location, with color denoting the visited location. A clear structural distinction emerges: some events form location-based clusters, while others are spatially mixed with no separation by destination. The embedding space thus loosely divides into an outer region of clustered behaviors and a central region of less location-dependent activity. 
This contrast suggests that the model distinguishes between different types of mobility behavior. We hypothesize that these structural differences reflect underlying activity types; for example, routine visits such as home or work (spatially consistent) versus exploratory or irregular leisure activities (spatially diffuse). Because the embedding space encodes patterns learned directly from event sequences, it provides a valuable basis for analyzing how behavioral variations relate to spatial semantics and for advancing context-aware modeling of mobility behavior.

\section{Role of behavioral and contextual factors}

\subsection{Identifying urban functions from POI features}\label{appendix:poi}

\begin{figure}[!htb]
  \centering
  \includegraphics[width=0.99\textwidth]{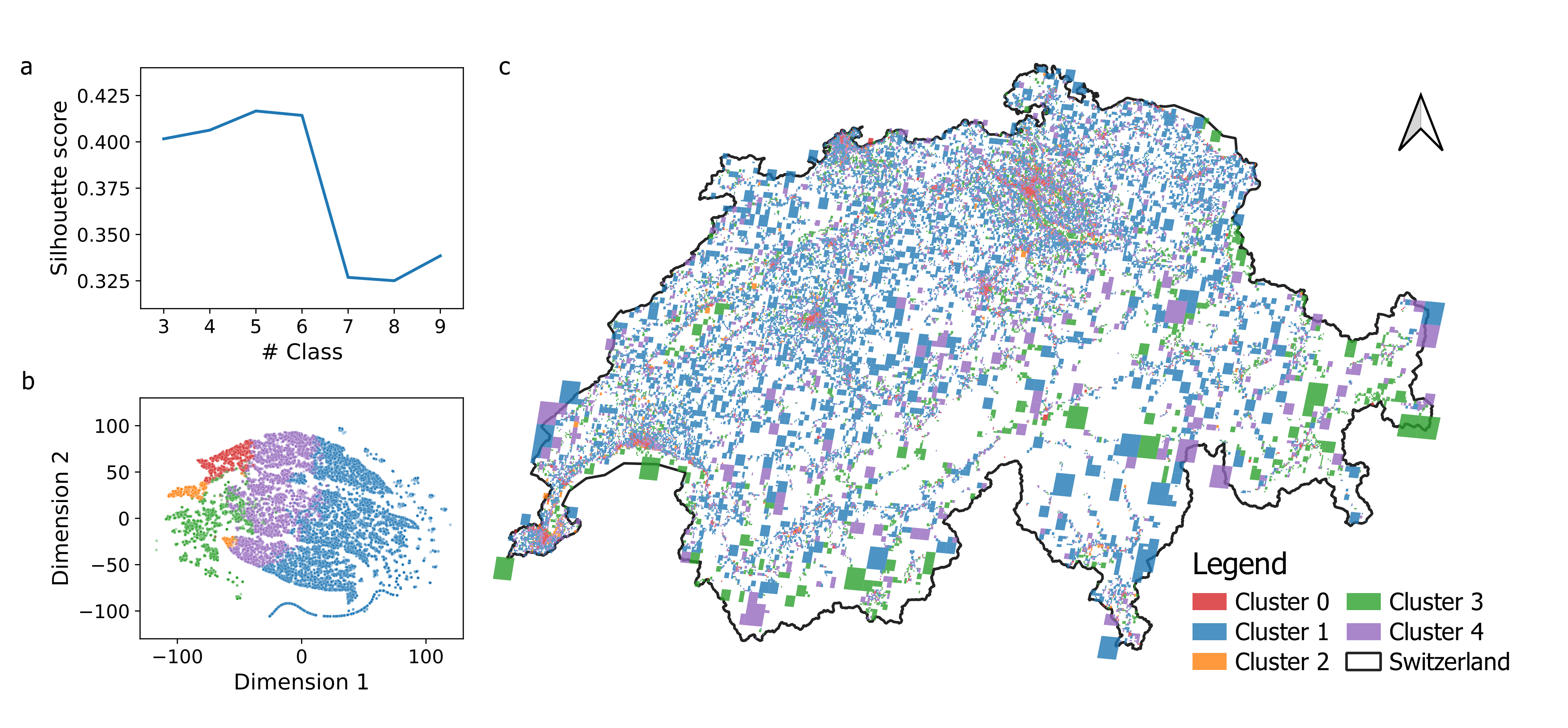}
  \caption{Identifying urban functions using location LDA vectors and $k$-means clustering.
  \textbf{a}. The silhouette score assesses clustering quality, with higher values indicating more compact and better-separated classes. The score suggests that five classes yield the best separation.
  \textbf{b}. Clustering results visualized using 2D t-SNE, where each point represents a location, color-coded by class.
  \textbf{c}. Spatial distribution of the identified location classes, using the same color scheme as in \textbf{b}.}
  
  \label{suppl_fig:poi_class}
\end{figure}

POIs reflect the socioeconomic activities in local regions and have been increasingly used to identify urban functions at a fine scale~\citep{yao_sensing_2017}. 
A well-established approach to functional identification involves obtaining a POI representation for each study unit, identifying cluster structures from these representations, and labeling the resulting classes based on expert knowledge~\citep{niu_delineating_2021}.
Here, we perform $k$-means clustering on the LDA functional descriptor of each location.
Using the silhouette score to assess clustering quality, we determine the optimal number of classes to be five (Figure~\ref{suppl_fig:poi_class}a).  
Visualizing the five-class clustering result using 2-dimensional t-SNE reveals that the data points within each class are compact, while those between classes are well-separated (Figure~\ref{suppl_fig:poi_class}b). 
To interpret the urban functions of each location group, we rely on the spatial distribution of classes (Figure~\ref{suppl_fig:poi_class}c) and the POI enrichment factor for each class (Table~\ref{tab:ei_poi}).  
The latter quantifies the relative occurrence of different categories of POIs. For each location class $i$ and POI category $q$, the enrichment factor $EF_{i}^{q}$ is defined as: 
\begin{equation}
    EF_{i}^{q} = \frac{N_{i}^{q}/N_{i}}{N^{q}/N}
\end{equation}
where \(N_{i}^{q}\) represents the number of category \(q\) POIs in location class \(i\), \(N_{i}\) denotes the total number of POIs in location class \(i\), \(N^{q}\) indicates the total number of category \(q\) POIs, and \(N\) signifies the total number of POIs in the study area.

We label Cluster 0 as the \textit{City center} due to its high prevalence in urban areas and the significant frequency of catering, health, and monetary-related POIs. 
Locations in Cluster 1 are dispersed outside the city and exhibit a high density of public facilities, leading to their classification as \textit{Residential \& Countryside}. 
Clusters 2 and 3 are designated as \textit{Leisure \& Recreation} and \textit{Nature \& Tourism}, respectively, due to their notable concentrations of leisure and nature-related POIs. 
Lastly, locations in Cluster 4 are typically situated near Cluster 0, and given their relatively even distribution of POIs, we classify this cluster as \textit{Suburbs \& Developing areas}.

The identification of urban functions is based on the locations' LDA functional descriptors, which are the same data as employed by \method. 
Since the process primarily relies on the structure of the data itself, we believe that this result offers a direct illustration of the information used in \method.

\begin{table}[!tb]
    \caption{POI enrichment across functional clusters. First-level POI categories with more than 1,000 instances in the study area were considered. For each cluster, the top three POI types with enrichment factors greater than 1 are highlighted in bold.}
    \label{tab:ei_poi}
    \centering
    \begin{tabular}{@{}cccccc@{}}
    \toprule 
    POI category               & Cluster 0   & Cluster 1 & Cluster 2 & Cluster 3 & Cluster 4 \\\midrule
    Accommodation   &  1.28       & 0.90      & 0.41          &  0.92         & 0.97\\
    Catering	    &  \textbf{1.88}       &  0.71     & 0.44          & 0.43          & 0.82           \\
    Health          &  \textbf{2.31}       & 0.46      & 0.12          & 0.04          & 0.85     \\
    Leisure         &  0.48      & \textbf{1.22}     & \textbf{3.26}          & 0.28          & \textbf{1.08}      \\
    Money         &  \textbf{2.21}      & 0.56     & 0.04          &   0.06        & 0.82      \\
    Natural         &  0.04      & 0.86     & 0.14          & \textbf{7.15}          & 0.87      \\
    Others         &  1.01      & 0.83     & 0.87          & \textbf{1.81}          & \textbf{1.03}      \\
    Place of worships         &  0.56      & \textbf{1.68}     & 0.22          & 0.32          & 0.85      \\
    Public         &  0.90      & \textbf{1.26}     & 0.66          & 0.13          & 0.99      \\
    Shopping         &  1.59      & 0.83     & 0.37          & 0.31          & 0.92      \\
    Tourism         &  0.54      & 1.05     & 0.55          & \textbf{3.08}          &  0.97     \\
    Traffic         &  1.00      & 1.09     & 0.64          & 0.50          &  \textbf{1.02}     \\
    
    \bottomrule
    \end{tabular}
    
\end{table}

\subsection{Effect of representing activity attributes and spatial context}\label{appendix:ablation}

\begin{table}[!b]
    \caption{Ablation study on activity attributes (location metrics). The full model is compared to reduced variants: one using location only (base), and others including time (base + time use) or travel mode (base + travel mode). Wasserstein distances are reported for number of visits per location (\( f_k \)), jump length (\( f_{\Delta r} \)), radius of gyration (\( f_{r_{u}} \)), uncorrelated entropy (\( S_{\text{unc}} \)), and temporal entropy (\( S_{\text{temp}} \)); best-performing values are shown in bold.}
    \label{tab:suppl_ab_location}
    \centering
    \begin{tabular}{@{}lccccc@{}}
    \toprule 
                                & $f_k$ & \( f_{\Delta r} \) & \( f_{r_{u}} \) & \( S_{\text{unc}} \) & $S_{\text{temp}}$ \\ \midrule
    Base                    & 0.649 {\footnotesize $\pm$0.003} & 3871 {\footnotesize $\pm$16} & 901 {\footnotesize $\pm$21} & 0.216 {\footnotesize $\pm$0.001} & 0.110 {\footnotesize $\pm$0.0004}            \\ 
    Base + time use         & 0.543 {\footnotesize $\pm$0.002} & \textbf{2280} {\footnotesize $\pm$21}  & 625 {\footnotesize $\pm$19} & 0.144 {\footnotesize $\pm$0.001} & 0.101 {\footnotesize $\pm$0.0005}           \\ 
    Base + travel mode      & 0.519 {\footnotesize $\pm$0.002} & 3493 {\footnotesize $\pm$15} & 804 {\footnotesize $\pm$13} & 0.136 {\footnotesize $\pm$0.001} & 0.088 {\footnotesize $\pm$0.0003}      \\ 
    Full model              & \textbf{0.262} {\footnotesize $\pm$0.004} & 2318 {\footnotesize $\pm$14} & \textbf{574} {\footnotesize $\pm$17} & \textbf{0.080} {\footnotesize $\pm$0.001} & \textbf{0.036} {\footnotesize $\pm$0.001}          \\  \bottomrule

    \end{tabular}
    
\end{table}

We conduct ablation studies on \method{} to assess the contributions of activity attributes and contextual information. Activity attributes serve dual roles in the model: they are embedded jointly with location data to support the learning of mobility behavior (Eq.~\ref{equation:attribute_embed}), and they provide additional supervisory signals through the loss function (Eq.~\ref{equation:combine_loss}). We begin with a model trained solely on location sequences and progressively incorporate time and travel mode attributes into the pipeline. All other training configurations are held constant to ensure a fair comparison.
The results are summarized in Table~\ref{tab:suppl_ab_location} and~\ref{tab:suppl_ab_activity}, where model performance is measured using the Wasserstein distance between generated and real sequences across a range of mobility metrics. Each model variant is simulated three times with different random seeds, and we report the mean and standard deviation of the resulting Wasserstein distances.
While the base model, trained only on historical location data, already performs strongly relative to classical mobility baselines (see performance numbers in Table~\ref{tab:distance_loc}), the inclusion of either time or travel mode data leads to further improvements. The full model, which incorporates all behavioral attributes, achieves the best overall performance, with a slight decline observed only in the jump length metric. These findings support our claim that jointly modeling behavioral attributes results in a more realistic reproduction of individual mobility patterns.

\addtolength{\tabcolsep}{-1pt}
\begin{table}[!tb]
    \caption{Ablation study on activity attributes (activity-travel metrics). The full model is compared to reduced variants: one incorporating only time (base + time use) and another incorporating only travel mode (base + travel mode). Wasserstein distances are reported for activity duration ($f_d$), activity start time ($f_{t}$), daily visited locations ($f_k^{\text{day}}$), unique daily visited locations ($f_{\left| k \right|}^{\text{day}}$), mobility motifs ($f_{\text{motifs}}$), and travel mode ($f_{m}$); best-performing values are shown in bold.}
    \label{tab:suppl_ab_activity}
    \centering
    \begin{tabular}{@{}lcccccc@{}}
    \toprule 
    & $f_d$ & $f_t$ & $f_k^{\text{day}}$ & $f_{\left| k \right|}^{\text{day}}$ & $f_{\text{motifs}}$ & $f_{m}$ \\ \midrule
    
    Base + time use         & 0.904 {\footnotesize $\pm$0.01} & 0.73 {\footnotesize $\pm$0.01} & 0.447 {\footnotesize $\pm$0.002} & 0.273 {\footnotesize $\pm$0.001} & 0.233 {\footnotesize $\pm$0.01} & -            \\ 
    Base + travel mode      & - & - & - & - & - & 0.113 {\footnotesize $\pm$0.001}        \\ 
    Full model              & \textbf{0.473} {\footnotesize $\pm$0.001} & \textbf{0.52} {\footnotesize $\pm$0.01} & \textbf{0.292} {\footnotesize $\pm$0.01} & \textbf{0.224} {\footnotesize $\pm$0.002} & \textbf{0.142} {\footnotesize $\pm$0.004} & \textbf{0.048} {\footnotesize $\pm$0.002} \\ \bottomrule

    \end{tabular}
    
\end{table}
\addtolength{\tabcolsep}{1pt}

Movement-related context, including location geometry and surrounding land-use functions, is integrated with activity attributes within the embedding space. We hypothesize that this integration enhances \method{}’s ability to characterize locations and is particularly useful for identifying novel locations for exploration. To evaluate this, we compare the full model with a variant that excludes contextual information. As described in the main text, generated locations are grouped into three sets based on their presence in the observed sequences: Set A includes locations common to both observed and generated data, while Sets B and C consist of locations unique to the observed and generated data, respectively. We then compute the Wasserstein distance to quantify discrepancies in the distribution of visits to functional areas. 
We expect the model with contextual information to generate more similar visitation patterns for novel locations (i.e., comparisons between Sets B and C), and more distinct patterns between shared and novel locations (i.e., comparisons between Sets A and C), than the model without contextual information. This expectation is supported by the Wasserstein distance results in Table~\ref{tab:suppl_ab_context}, although the differences between models for Sets B and C are not statistically significant. This is likely due to the relatively low visit frequencies to these locations, which result in greater variability in visitation patterns.

\begin{table}[!htbp]
    \caption{Ablation study on contextual information. The full model is compared to a variant without contextual inputs (\method{} w/o ctx). Locations are grouped based on their presence in real and generated sequences. Wasserstein distances quantify differences in the distribution of visits to functional areas, including Set $A$ (real) vs. Set $A$ (generated) ($d_{A}$), Set C vs. Set B ($d^C_{B}$), Set C vs. Set $A$ (real) ($d^C_{A_{observe}}$), and Set C vs. Set $A$ (generated) ($d^C_{A_{generate}}$). Arrows indicate whether smaller or larger values are preferred.}
    \label{tab:suppl_ab_context}
    \centering
    \begin{tabular}{@{}lcccc@{}}
    \toprule 
                           & $d_{A}$ $\downarrow$ & $d^C_{B}$ $\downarrow$ & $d^C_{A_{observe}}$ $\uparrow$ & $d^C_{A_{generate}}$ $\uparrow$ \\ \midrule
    \method{} w/o ctx      & 0.025 {\footnotesize $\pm$0.001}  & 0.153 {\footnotesize $\pm$0.067} & 0.548 {\footnotesize $\pm$0.001} & 0.545 {\footnotesize $\pm$0.001}\\ 
    \method{}              & 0.009 {\footnotesize $\pm$0.001}  & 0.105 {\footnotesize $\pm$0.006} & 0.560 {\footnotesize $\pm$0.005} & 0.558 {\footnotesize $\pm$0.003}\\ \bottomrule
    \end{tabular}

\end{table}

\subsection{Comparison with mobility flow prediction models}\label{appendix:flow}

We benchmark \method{} against established flow models to evaluate how effectively it leverages contextual information when predicting mobility flows in unseen locations. 
We adopt the flow prediction task with the objective to estimate the flow $y(l_i,l_j)$ between each origin-destination (OD) pair $l_i$ and $l_j$ per unit time, given the total outflow $O_i$ from location $l_i$~\citep{simini_deep_2021}. 
That is, the task involves learning to distribute the total outflow from a given origin across all potential destinations based on location characteristics. 
Typical features include inter-location distance, population density, and contextual variables such as land use and POI types, which reflect the activity options available at each location.
These correspond closely to the contextual inputs incorporated in \method{}.

To ensure consistency in temporal coverage and enable a fair comparison, we construct OD flow pairs from the test set of mobility event sequences. 
Locations are partitioned into two non-overlapping subsets: (i) $\mathbb{S}_{\text{seen}}$, comprising locations observed during the training period, and (ii) $\mathbb{S}_{\text{unseen}}$, containing locations that appear only during testing.
Flow models are trained on OD pairs originating from $\mathbb{S}_{\text{seen}}$ and evaluated on flows from $\mathbb{S}_{\text{unseen}}$, providing a direct assessment of generalization to novel locations. 
For \method{}, we aggregate simulated event sequences into OD flow pairs and evaluate performance exclusively on flows originating from $\mathbb{S}_{\text{unseen}}$.
We next describe the flow models and feature sets used for comparison:

\begin{itemize}
    \item Gravity~\citep{zipf_p1_1946}. 
    We implement the singly constrained gravity model, which estimates flow $\hat{y}(l_i,l_j)$ from origin $l_i$ to destination $l_j$ based on population sizes $m_i$, $m_j$, and the distance $d_{ij}$:
    \begin{equation}
        \hat{y}(l_i,l_j) = O_i P_{ij} = O_i\frac{m_j^{\beta_1}f(d_{ij})}{\sum_k m_k^{\beta_1}f(d_{ik})}
    \end{equation}
    Here, $P_{ij}$ denotes the normalized probability of travel from $l_i$ to $l_j$, and $f(\cdot)$ is the deterrence function, specified as either exponential, $f(d) = \exp(\beta_2 d)$, or power-law, $f(d) = d^{\beta_2}$. 
    The model parameters $\beta_1$ and $\beta_2$ are fitted to the training flows using maximum likelihood estimation. 
    Preliminary experiments suggest slightly better performance using the power-law form.
    \item Deep Gravity~\citep{simini_deep_2021}. 
    This model generalizes the classical gravity framework by introducing hidden layers and nonlinearities, effectively converting it into a feedforward neural network.
    It incorporates additional geographic context for each location, including land use types, road network lengths, and counts of facilities related to transport, food, health, education, and retail. 
    Each OD pair is described using 39 input features: 18 contextual attributes for both origin and destination, the distance between them, and their population sizes. 
    We follow the original implementation, using the same feature set, architecture, and training procedure.
    \item Random Forest. 
    We use the same 39 input features as Deep Gravity to train a random forest regressor for predicting OD flow intensities.
    We employ 1000 estimators~\citep{cabanas_human_2025} and the default hyperparameters from scikit-learn~\citep{scikit_learn}.
    
\end{itemize}

\begin{table}[!t]
    \caption{Reconstructing mobility flows at novel locations. Simulated flows are compared with real flows at locations not seen during training. For \method{}, individual sequences are aggregated into an origin-destination matrix to enable comparison with flow-based models. Evaluation metrics include the common part of commuters (CPC), Pearson correlation coefficient (Pearson r), and Jensen-Shannon divergence (JSD). Arrows indicate the preferred direction for each metric. Entropy quantifies the spatial dispersion of visits to novel locations. Best-performing values are shown in bold.}
    \label{tab:flow_all}
    \centering
    \begin{tabular}{@{}lccccc@{}}
    \toprule 
                        & CPC $\uparrow$   & Pearson r $\uparrow$ & JSD $\downarrow$  & Entropy \\\midrule
    Gravity power-law   &  0.02  & 0.19   & 0.81  & 9.88    \\
    Random Forest       &  0.02  & 0.17   & 0.79  & 8.75    \\
    Deep Gravity        &  0.08  & 0.25   & 0.77  & 8.33    \\ 
    \method{} (Ours)    &  \textbf{0.22}  & \textbf{0.47}   & \textbf{0.69}  & \textbf{7.16}    \\ \midrule 
    Data                &        &        &       & 6.92    \\ \bottomrule
    \end{tabular}
    
\end{table}

We evaluate model performance on flows originating from $\mathbb{S}_{\text{unseen}}$ using established mobility flow metrics: common part of commuters, Pearson correlation coefficient, and Jensen-Shannon divergence. 
Summary results are presented in Table~\ref{tab:flow_all}.
To further assess spatial distributions, we compute the entropy of disaggregated destination probabilities. 
These are compared with empirical distributions and those generated by \method{}.
Across all metrics, \method{} consistently achieves the highest performance and most closely reproduces the observed spatial distribution of location visits, outperforming all baseline models.
We attribute these gains to the integration of detailed spatial features (e.g., urban function distributions via LDA) alongside the joint modeling of contextual and behavioral factors.

\section{Sensitivity and generalization analyses}

\subsection{Sensitivity to sequence length}\label{appendix:sequence}
To assess whether \method{}'s performance depends on the chosen sequence lengths, we conduct sensitivity analyses for both the input traveled sequence length $k_2$ and the generated target length $k_1$. For sensitivity to $k_2$, the generated target horizon is fixed at 50 future activity events while the number of observed input events is varied. For sensitivity to $k_1$, both generated and empirical target sequences are capped to the same number of future events. In both analyses, we report the corresponding calendar days covered by each event horizon to clarify the effective temporal extent of each setting.

Longer input histories generally improve generation quality (Table~\ref{tab:src_sensitivity}). Increasing $k_2$ from 2 to 50 events substantially reduces the Wasserstein distances across most location and activity-travel metrics, indicating that additional observed history helps the model recover individual spatial preferences and temporal regularities. The original variable-length traveled setting performs best or close to best for most metrics, suggesting that the model benefits from preserving longer observed histories whenever they are available, and that sufficient input helps infer individual preferences more reliably. This motivates future work on more compact representations of individual preferences, such as learned user embeddings or summary statistics, that could maintain generation quality with shorter observed inputs.

\addtolength{\tabcolsep}{-1pt}

\begin{table}[!hb]
    \caption{Sensitivity of \method{} performance to the input traveled length \(k_2\). Wasserstein distances are reported for location and activity-travel metrics; lower values indicate better alignment with empirical target sequences. The target horizon is fixed to 50 events, and days covered by the input traveled sequence are reported as mean \(\pm\) SD. ``Variable'' denotes the original variable-length traveled sequence used in the main evaluation.}
    \label{tab:src_sensitivity}
    \centering
    \begin{tabular}{@{}lcccccc cccccc@{}}
    \toprule
    \multirow{2}{*}{$k_2$}
    & \multirow{2}{*}{Days covered}
    & \multicolumn{5}{c}{Location metrics}
    & \multicolumn{6}{c}{Activity-travel metrics} \\
    \cmidrule(lr){3-7} \cmidrule(lr){8-13}
    & 
    & $f_k$
    & $f_{\Delta r}$
    & $f_{r_u}$
    & $S_{\mathrm{unc}}$
    & $S_{\mathrm{temp}}$
    & $f_d$
    & $f_t$
    & $f_k^{\mathrm{day}}$
    & $f_{|k|}^{\mathrm{day}}$
    & $f_{\mathrm{motifs}}$
    & $f_m$ \\
    \midrule
    2   & $0.5$ {\footnotesize $\pm$0.4}  & 1.27 & 27567 & 7107 & 1.00 & 0.72 & 0.72 & 1.57 & 0.68 & 1.08 & 0.89 & 0.17 \\
    5   & $1.2$ {\footnotesize $\pm$0.8}  & 0.63 & 15457 & 5680 & 0.42 & 0.28 & 0.37 & 1.55 & 0.44 & 0.62 & 0.52 & 0.13 \\
    20  & $4.6$ {\footnotesize $\pm$2.0}  & 0.32 & 5146  & 1797 & 0.14 & 0.06 & 0.36 & 0.69 & 0.41 & 0.29 & 0.18 & 0.08 \\
    50  & $12.2$ {\footnotesize $\pm$3.9} & 0.32 & 3171  & 647  & 0.12 & 0.05 & 0.41 & 0.65 & 0.35 & 0.25 & 0.14 & 0.08 \\
    Variable & $19.9$ {\footnotesize $\pm$2.6} & 0.26 & 2318 & 575 & 0.08 & 0.04 & 0.47 & 0.52 & 0.29 & 0.22 & 0.14 & 0.05 \\
    \bottomrule
    \end{tabular}
    
\end{table}

The 50-event horizon provides a stable multi-day evaluation setting (Table~\ref{tab:tgt_sensitivity}). At very short horizons, some location metrics yield deceptively small distances because only limited mobility diversity is expressed, while sequence-based metrics such as temporal entropy and motifs become undefined or unstable. At longer horizons, generation remains feasible, but displacement and radius of gyration measures show larger deviations at 100 events, suggesting that maintaining spatial consistency becomes more challenging as the generated horizon increases. Future work could investigate generation strategies that better support longer horizons, for example by training on longer sequences or generating in successive segments.

\begin{table}[!th]
    \caption{Sensitivity of \method{} performance to the generated target length \(k_1\). Wasserstein distances are reported for location and activity-travel metrics; lower values indicate better alignment with empirical target sequences capped to the same number of future events. The 50-event horizon is the default setting, and days covered by the target sequence are reported as mean \(\pm\) SD. ``--'' indicates metrics that are undefined or unstable for very short event sequences.}
    \label{tab:tgt_sensitivity}
    \centering
    \begin{tabular}{@{}lcccccc cccccc@{}}
    \toprule
    \multirow{2}{*}{$k_1$}
    & \multirow{2}{*}{Days covered}
    & \multicolumn{5}{c}{Location metrics}
    & \multicolumn{6}{c}{Activity-travel metrics} \\
    \cmidrule(lr){3-7} \cmidrule(lr){8-13}
    & 
    & $f_k$
    & $f_{\Delta r}$
    & $f_{r_u}$
    & $S_{\mathrm{unc}}$
    & $S_{\mathrm{temp}}$
    & $f_d$
    & $f_t$
    & $f_k^{\mathrm{day}}$
    & $f_{|k|}^{\mathrm{day}}$
    & $f_{\mathrm{motifs}}$
    & $f_m$ \\
    \midrule
    2   & $0.8$ {\footnotesize $\pm$0.6}  & 0.05 & 2736 & 1440 & 0.07 & --   & 4.72 & 4.05 & 0.28 & 0.19 & --   & 0.16 \\
    5   & $1.7$ {\footnotesize $\pm$1.0}  & 0.01 & 2437 & 768  & 0.01 & 0.02 & 2.70 & 2.69 & 0.56 & 0.32 & 0.05 & 0.15 \\
    20  & $4.8$ {\footnotesize $\pm$2.0}  & 0.20 & 2428 & 539  & 0.09 & 0.06 & 0.35 & 0.71 & 0.49 & 0.34 & 0.13 & 0.11 \\
    50  & $12.7$ {\footnotesize $\pm$4.0} & 0.26 & 2318 & 575  & 0.08 & 0.04 & 0.47 & 0.52 & 0.29 & 0.22 & 0.14 & 0.05 \\
    100 & $20.5$ {\footnotesize $\pm$7.3} & 0.15 & 4050 & 1533 & 0.03 & 0.02 & 0.60 & 0.40 & 0.27 & 0.17 & 0.14 & 0.10 \\
    \bottomrule
    \end{tabular}
    
\end{table}

\subsection{Generalization to unseen users}\label{appendix:user_split}

The main evaluation uses a temporal split over users who appear in both the training and test sets. This setting evaluates how well the model continues held-out future trajectories, but performance may partly reflect extrapolation from previously observed individual routines. To assess whether \method{} captures mobility structure beyond known individuals, we conduct a held-out-individual experiment in which 150 users are reserved for validation and 150 for testing, with their mobility trajectories removed from the training set. The resulting evaluation contains two regimes. The seen-user setting evaluates held-out future days of training users, as in the main evaluation, while the unseen-user setting uses trajectories from users entirely absent from training. In both regimes, the model is conditioned on the source observation window and generates the next 50 activity-travel events.

A comparison between seen and unseen users must account for coverage of the discrete location space. \method{} learns representations for S2 grid cells from locations observed during training; cells absent from the training trajectories receive no positive training examples. Limited coverage can therefore degrade generation quality in addition to the challenge of transferring to a new individual. For each user, we compute source-location coverage as the fraction of unique locations in their source sequence that appeared in the training data; the mean coverage among seen users is $94.8\%$. We accordingly divide the unseen users at $95\%$ coverage into a high-coverage subgroup ($n=58$), which most closely matches the seen-user setting in embedding support, and a lower-coverage subgroup ($n=92$), which tests robustness when a smaller proportion of the user's source locations was observed during training. Together, the two subgroups comprise all 150 held-out users.

Baseline models are evaluated under the same information constraint as \method{}. Each is conditioned only on the source sequence available at test time and generates the next 50 events. For mechanistic baselines, individual parameters are estimated from the source sequence, while population-level priors are fitted exclusively on the seen training users. This protocol prevents any baseline from accessing information unavailable to \method{}.

Table~\ref{tab:unseen_baseline_comparison_coverage} reports the results across the two coverage subgroups. Among the baselines, the learning-based location-only models perform poorly for unseen users, whereas several mechanistic baselines retain comparatively stable spatial performance across coverage levels. In the high-coverage subgroup, \method{} remains close to the seen-user reference on visitation frequency and uncorrelated entropy and outperforms DITRAS and TimeGeo across all comparable activity-travel metrics. At the same time, larger differences remain for displacement and radius of gyration, suggesting that measures of spatial extent are sensitive to the shift from seen to unseen users.

In the lower-coverage subgroup, the location metrics deteriorate further, particularly the distance-based measures, and the strongest mechanistic baselines, Container and DITRAS, achieve lower errors across all location metrics. Their smaller changes across coverage subgroups suggest that the deterioration of \method{} is partly associated with its dependence on learned location representations. However, because the two subgroups may also differ in their underlying mobility characteristics, the effect cannot be attributed exclusively to embedding coverage. By contrast, \method{} continues to outperform DITRAS and TimeGeo across all comparable activity-travel metrics. Activity-travel regularities therefore appear to transfer more robustly than spatial usage patterns.

\begin{table*}[!t]
    \caption{Comparison between \method{} and baseline models for unseen users with high ($\geq 95\%$) and lower ($<95\%$) source-location coverage. Together, the two subgroups comprise all 150 held-out users; the seen-user continuation setting is included as a reference. Wasserstein distances are reported for location and activity-travel metrics. ``--'' indicates metrics not defined for baselines that do not generate the corresponding attribute.}
    \label{tab:unseen_baseline_comparison_coverage}
    \centering
    \begin{tabular}{@{}lccccc cccccc@{}}
    \toprule
    \multirow{2}{*}{Method}
    & \multicolumn{5}{c}{Location metrics}
    & \multicolumn{6}{c}{Activity-travel metrics} \\
    \cmidrule(lr){2-6} \cmidrule(lr){7-12}
    & $f_k$
    & $f_{\Delta r}$
    & $f_{r_u}$
    & $S_{\mathrm{unc}}$
    & $S_{\mathrm{temp}}$
    & $f_d$
    & $f_t$
    & $f_k^{\mathrm{day}}$
    & $f_{|k|}^{\mathrm{day}}$
    & $f_{\mathrm{motifs}}$
    & $f_m$ \\
    \midrule

    \multicolumn{12}{@{}l}{\textit{High source-location coverage ($\geq95\%$, $n=58$ users)}} \\
    \addlinespace
    EPR
    & 1.15 & 9204 & 5671 & 0.45 & 0.35 & -- & -- & -- & -- & -- & -- \\
    Container
    & 0.58 & 3055 & 1619 & 0.19 & 0.20 & -- & -- & -- & -- & -- & -- \\
    Markov
    & 1.45 & 1607 & 8789 & 0.83 & 0.76 & -- & -- & -- & -- & -- & -- \\
    MHSA
    & 1.51 & 12800 & 14102 & 0.31 & 0.26 & -- & -- & -- & -- & -- & -- \\
    MoveSim
    & 7.21 & 18063 & 22255 & 1.01 & 0.90 & -- & -- & -- & -- & -- & -- \\
    \addlinespace
    DITRAS
    & 0.46 & 9740 & 5615 & 0.14 & 0.15
    & 1.94 & 1.72 & 0.70 & 0.27 & 0.50 & -- \\
    TimeGeo
    & 1.12 & 13650 & 10487 & 0.39 & 0.41
    & 2.89 & 1.98 & 1.84 & 1.57 & 1.54 & -- \\
    \method{}
    & 0.67 & 7048 & 5375 & 0.15 & 0.21
    & 0.35 & 0.94 & 0.19 & 0.11 & 0.17 & 0.29 \\

    \midrule
    \multicolumn{12}{@{}l}{\textit{Lower source-location coverage ($<95\%$, $n=92$ users)}} \\
    \addlinespace
    EPR
    & 1.70 & 9876 & 5998 & 0.47 & 0.37
    & -- & -- & -- & -- & -- & -- \\
    Container
    & 0.49 & 4789 & 2202 & 0.11 & 0.11
    & -- & -- & -- & -- & -- & -- \\
    Markov
    & 2.14 & 1828 & 9608 & 0.88 & 0.80
    & -- & -- & -- & -- & -- & -- \\
    MHSA
    & 2.14 & 17112 & 18661 & 0.39 & 0.28
    & -- & -- & -- & -- & -- & -- \\
    MoveSim
    & 9.27 & 38692 & 39841 & 1.17 & 0.95
    & -- & -- & -- & -- & -- & -- \\
    \addlinespace
    DITRAS
    & 0.55 & 11202 & 6811 & 0.15 & 0.17
    & 2.10 & 1.41 & 0.81 & 0.31 & 0.41 & -- \\
    TimeGeo
    & 2.11 & 15015 & 11532 & 0.52 & 0.44
    & 2.48 & 1.57 & 1.56 & 1.68 & 1.46 & -- \\
    \method{}
    & 1.32 & 20363 & 12687 & 0.32 & 0.31
    & 0.92 & 1.15 & 0.32 & 0.18 & 0.28 & 0.08 \\
    \midrule
    \multicolumn{12}{@{}l}{\textit{Seen-user continuation reference}} \\
    \addlinespace
    \method{} (seen users)
    & 0.65 & 3236 & 1146 & 0.18 & 0.12
    & 0.78 & 0.46 & 0.45 & 0.29 & 0.26 & 0.09 \\

    \bottomrule
    \end{tabular}
\end{table*}

\addtolength{\tabcolsep}{1pt}

Overall, the held-out-individual experiment provides a stronger test of \method{}'s generative capability and supports two conclusions. First, the model captures population-level activity-travel regularities beyond the continuation of known individual routines. Second, its generalization to unseen users remains limited for spatial behavior and deteriorates substantially when source-location coverage is lower. The gap between the seen-user reference and the high-coverage subgroup suggests that performance in the main setting benefits from user-specific information learned during training, while the additional deterioration in the lower-coverage subgroup is consistent with sensitivity to missing embedding support. These effects point to complementary directions for future work. Pretrained geographic representations could provide informative features for cells absent from the training trajectories, while conditioning on explicit user attributes or representations learned from the source observations could improve adaptation to individual variation.

\section{Deriving experienced income segregation}\label{appendix:segregation}

\begin{figure}[!htb]
  \centering
  \includegraphics[width=0.7\textwidth]{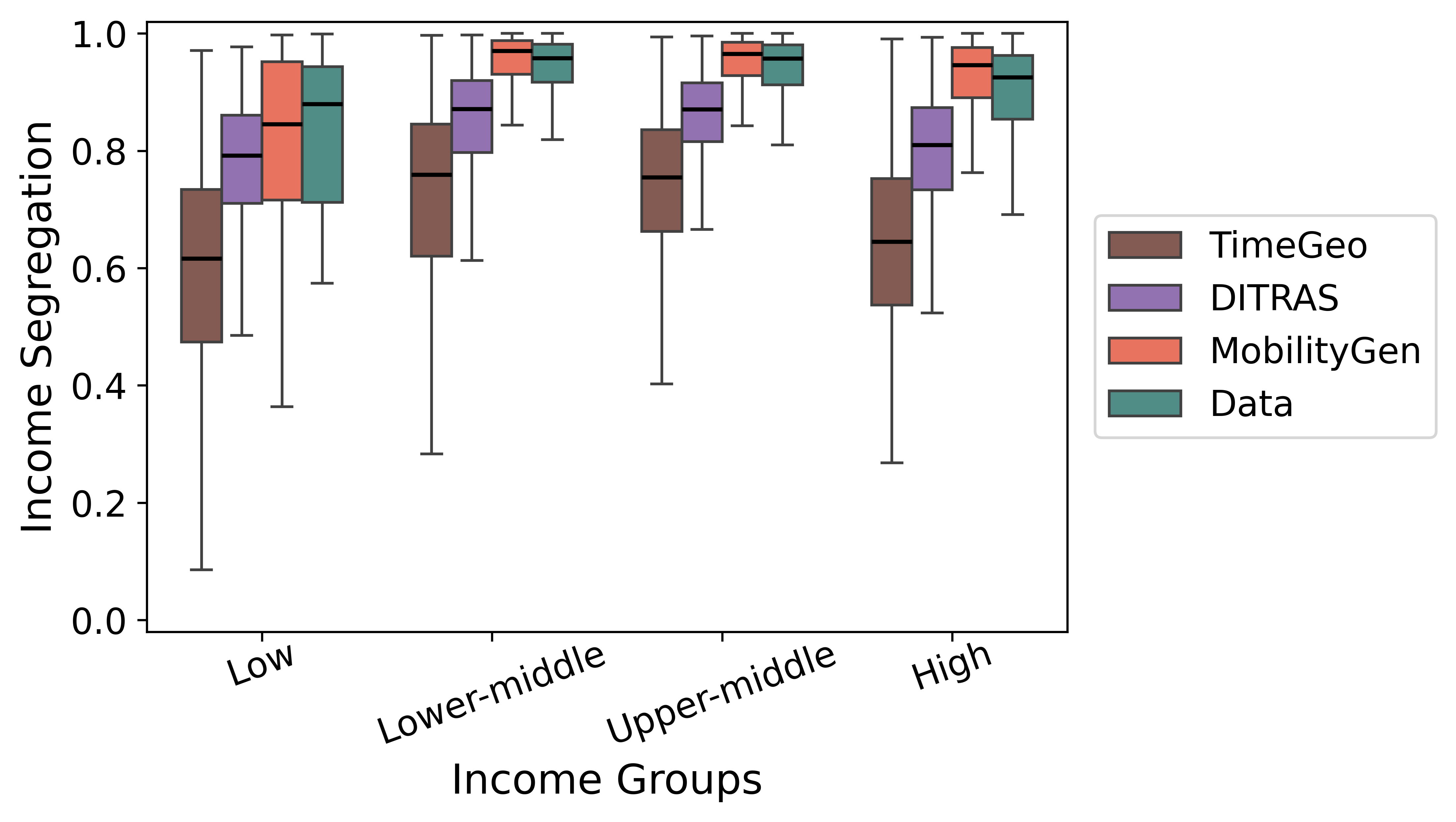}
  \caption{Variation in income segregation across groups. Boxplots of individual-level income segregation scores for four income groups (low, lower-middle, upper-middle, and high), comparing observed data (teal) with sequences generated by TimeGeo (brown), DITRAS (purple), and \method{} (red). \method{} closely reproduces the empirical distributions and outperforms the baseline simulators in capturing variation in segregation levels across groups.}
  \label{suppl_fig:segregation}
\end{figure}

To measure income segregation, we divide individuals into four roughly equal-sized quartiles according to their reported monthly household income: low ($< 4,000$ CHF), lower-middle ($4,000-8,000$ CHF), upper-middle ($8,000-12,000$ CHF), and high ($> 12,000$ CHF). We then apply the experienced income segregation framework~\citep{moro_mobility_2021}, described in detail below.

At the location level, segregation is computed from the proportion of time individuals in income quartile $q$ spend at each location $l$, denoted by $\tau_{ql}$. Assuming that a fully integrated location corresponds to an equal distribution of time spent across quartiles (i.e., $\tau_{ql} = \frac{1}{4}$), the location-level segregation score $S_l$ is defined as: $S_{l}=\frac{2}{3}\sum_{q}\left| \tau_{ql} - \frac{1}{4} \right|$.
This score, ranging from 0 (perfect integration) to 1 (complete segregation), quantifies the imbalance in visitation by income groups.

To measure segregation at the individual level, we calculate each person's relative exposure to income quartiles based on the locations they visit. Let $\tau_{ul}$ represent the proportion of time individual $u$ spends at location $l$. Their exposure to quartile $q$, denoted by $\tau_{uq}$, is then: $\tau_{uq}=\sum_l\tau_{ul}\tau_{ql}$. Using this, the individual-level segregation score $S_u$ is defined analogously to the location-level score:
\begin{equation}
    S_{u}=\frac{2}{3}\sum_{q}\left| \tau_{uq} - \frac{1}{4} \right|
\end{equation}

As with $S_l$, a score of $S_u = 0$ indicates perfectly balanced exposure across income groups, while $S_u = 1$ reflects exclusive exposure to a single group.

This formulation approximates co-presence by averaging the time spent at each location across income groups, rather than relying on exact co-occurrence events. As such events are extremely sparse in real mobility data, this time-invariant approach provides a practical way to estimate individual-level interaction patterns. We compute individual-level income segregation scores for each real and simulated event sequence in the testing set. Since multiple sequences may originate from the same individual (and thus the same income group), and often involve repeated visits to similar locations, the resulting scores do not reflect population-wide segregation within the study area. Instead, they serve as a validation metric to assess how well \method{} reproduces the co-presence patterns observed in the real data.

\bibliography{mybibfile}

\clearpage


\end{document}